\documentclass[acmsmall,nonacm,preprint]{acmart}

\AtBeginDocument{%
  \providecommand\BibTeX{{%
    \normalfont B\kern-0.5em{\scshape i\kern-0.25em b}\kern-0.8em\TeX}}}

\setcopyright{acmcopyright}
\copyrightyear{2018}
\acmYear{2018}
\acmDOI{10.1145/1122445.1122456}

\acmJournal{JACM}
\acmVolume{37}
\acmNumber{4}
\acmMonth{8}

\renewcommand\footnotetextcopyrightpermission[1]{} % removes footnote with conference information in first column
\usepackage{lscape}
\usepackage[utf8]{inputenc}
\usepackage{makecell}
\usepackage{tcolorbox} 
\tcbuselibrary{listings,breakable}
\usepackage{graphicx}
\usepackage{array}
\usepackage{longtable}
\usepackage{comment}
\usepackage[font=small,skip=0pt]{caption}
\usepackage{enumitem}
\usepackage{booktabs}
\usepackage{hyperref}
\usepackage{float}
\usepackage{caption}
\usepackage{subcaption}
\usepackage{amsmath}
\usepackage{mathpazo}
\usepackage{datetime}
\usepackage{color, colortbl}

\definecolor{VeryHigh}{rgb}{0.90,0.44,0.43}
\definecolor{High}{rgb}{0.92,0.59,0.47}
\definecolor{Medium}{rgb}{0.94,0.72,0.51}
\definecolor{Low}{rgb}{0.68,0.81,0.52}
\definecolor{VeryLow}{rgb}{0.47,0.73,0.50}

\definecolor{Grey}{rgb}{0.85,0.85,0.85}

\usepackage{hyperref}
\hypersetup{
    colorlinks=true,
    linkcolor=blue,
    filecolor=magenta,      
    urlcolor=cyan,
    pdftitle={Overleaf Example},
    pdfpagemode=FullScreen,
    }

\usepackage{bookmark}
\usepackage{multirow}
\usepackage{xcolor}
\usepackage{color, colortbl}
\usepackage{tabularx}
\usepackage{algorithm,algorithmicx,algpseudocode}

\usepackage{url}
\usepackage{todonotes}
\usepackage{listings}
\input{prolog-language}

\makeatletter
\newcommand{\thickhline}{%
    \noalign {\ifnum 0=`}\fi \hrule height 1pt
    \futurelet \reserved@a \@xhline}
\newcolumntype{"}{@{\hskip\tabcolsep\vrule width 1pt\hskip\tabcolsep}}
\makeatother

\begin{document}

\title{Evaluating the Cybersecurity Risk of Real World, Machine Learning Production Systems}

\author{Ron Bitton}
\email{ronbit@post.bgu.ac.il}
\author{Nadav Maman}
\email{nadavmam@post.bgu.ac.il}
\author{Yuval Elovici}
\email{elovici@bgu.ac.il}
\author{Asaf Shabtai}
\email{ronbit@post.bgu.ac.il}
\affiliation{%
  \institution{Ben-Gurion University of the Negev}
%  \streetaddress{P.O. Box 1212}
  \city{Beer-Sheba}
  \country{Israel}
}

\author{Inderjeet Singh}
\email{inderjeet78@nec.com}
\author{Satoru Momiyama}
\email{satoru-momiyama@nec.com}
\affiliation{%
  \institution{NEC}
%  \streetaddress{1 Th{\o}rv{\"a}ld Circle}
  \country{Japan}}

\renewcommand{\shortauthors}{Bitton, et al.}

\begin{abstract}
Although cyberattacks on machine learning (ML) production systems can be harmful, today, security practitioners are ill equipped, lacking methodologies and tactical tools that would allow them to analyze the security risks of their ML-based systems.
In this SoK paper, we performed a comprehensive threat analysis of ML production systems.
In this analysis, we follow the ontology presented by NIST for evaluating enterprise network security risk and apply it to ML-based production systems.
Specifically, we (1) enumerate the assets of a typical ML production system, (2) describe the threat model (i.e., potential adversaries, their capabilities, and their main goal), (3) identify the various threats to ML systems, and (4) review a large number of attacks, demonstrated in previous studies, which can realize these threats. 
In addition, to quantify the risk of adversarial machine learning (AML) threat, we introduce a novel scoring system, which assign a severity score to different AML attacks. 
The proposed scoring system utilizes the analytic hierarchy process (AHP) for ranking, with the assistance of security experts, various attributes of the attacks. 
Finally, we developed an extension to the MulVAL attack graph generation and analysis framework to incorporate cyberattacks on ML production systems. 
Using the extension, security practitioners can apply attack graph analysis methods in environments that include ML components; thus, providing security practitioners with a methodological and  practical tool for evaluating the impact and quantifying the risk of a cyberattack targeting an ML production system.
\end{abstract}






\maketitle

\section{Introduction}
In the past few years, we have witnessed a revolution in the use and deployment of artificial intelligence (AI) and machine learning (ML) systems by organizations and large enterprises.
Gartner's 2019 CIO Survey shows that 37\% of enterprises have implemented some form of ML, which represents a 270\% increase in four years.\footnote{\scriptsize\url{https://www.gartner.com/en/newsroom/press-releases/2019-01-21-gartner-survey-shows-37-percent-of-organizations-have}} 
Furthermore, among Fortune 1,000 companies, 91.5\% of businesses have reported ongoing investment in ML-based technologies.\footnote{\scriptsize\url{http://newvantage.com/wp-content/uploads/2020/01/NewVantage-Partners-Big-Data-and-AI-Executive-Survey-2020-1.pdf}}
ML systems are utilized by enterprises in various use cases, including cybersecurity, fraud detection, financial trading, personalized marketing, resource optimization and allocation, healthcare, and autonomous vehicles.
In most of these use cases, critical decisions are made based on the processing and output of the ML system, thus a failure in these systems may have significant impact on business continuity, revenue, and, in some use cases, even human lives.

The increasing popularity of ML systems, coupled with their critical role in decision-making procedures, makes them an ideal target for attackers.
Unfortunately, ML systems are not vulnerability-free.
Like any software or product, these systems often include bugs that can be exploited by malicious actors.
ML systems also suffer from a special type of logical vulnerabilities stemming from inherent limitations of the underlying ML algorithms.
To exploit these vulnerabilities, attackers utilize an attack technique referred to as adversarial ML, which has already been demonstrated in security and ML research~\cite{carlini2017adversarial,szegedy2013intriguing,huang2011adversarial}.
Despite the fact that a cyberattack on ML systems can cause harm to systems, enterprises, and the people dependent on them; industry practitioners today are not equipped with the tactical and strategic tools needed to analyze, detect, protect, and respond to cyberattacks on their ML systems~\cite{kumar2020adversarial}.
Specifically, current risk assessment tools suffer from two main limitations.

\noindent \textbf{(i) Assessment granularity level.} Existing risk assessment tools represent the target environment at a very low resolution.
For example, in the MulVAL attack graph generation and analysis framework~\cite{ou2005mulval}, hosts are represented by their installed software, running services, file permissions, and network reachability. 
However, the business logic of these applications or services (such as in the case of ML applications) is not considered.
For example, the impact of manipulating an arbitrary file is not the same as manipulating a model's training data, since by manipulating a model's training data, the attacker can influence the model's decisions and by that, the attacker can extend the scope of the attack behind the specific host storing the training file.

\noindent \textbf{(ii) Adversarial ML.} Current risk assessment tools are aimed at modeling software and network vulnerabilities and accordingly consider traditional attack techniques.
However, ML systems also suffer from logical vulnerabilities, which can be exploited using adversarial ML.
Recent works have mostly focused in developing frameworks and libraries for evaluating the robustness of ML models to adversarial ML. 
Notable frameworks are the CleverHans adversarial examples library~\cite{papernot2018cleverhans}, the Adversarial Robustness Toolbox (ART)~\cite{nicolae2018adversarial}, the Foolbox toolbox~\cite{rauber2017foolbox}, the SecML library~\cite{melis2019secml}, and MLsploit~\cite{downing2019mlsploit}.
Although these frameworks, which implement various adversarial ML attack techniques, can be used to quantify the resilience of an ML model to different set of attack methods, they are focusing solely on implementing algorithms for finding adversarial examples, and measuring their performance against the target ML model.
However, these frameworks are not designed to quantify the risk of ML-based system to cybersecurity threats in general, and to adversarial ML threats in particular since they ignore the specific deployment of the target environment, as well as the different attributes of the attack technique (e.g., attacker model, attack impact, and attack performance), which are mandatory for practical risk assessment. 

In this paper, we take a significant step toward securing ML production systems by integrating these unique systems and their vulnerabilities into cybersecurity risk assessment frameworks.
We begin by exploring ML production systems and performing a comprehensive threat analysis, based on the ontology presented by NIST for evaluating an enterprise network security risk.
Specifically, we enumerate the main components/assets of ML production systems, describe the threat model (i.e., potential adversaries, their capabilities, and their main goal), identify the threats to ML systems, and review a large number of attacks, demonstrated in previous studies, which can realize the threats.

In addition, to quantifying the risk of adversarial machine learning (AML) threat, we introduce a novel scoring system, which assigns a severity score to different AML attacks.
The proposed scoring system utilizes the analytic hierarchy process (AHP) for ranking~\cite{saaty2008decision}, with the assistance of security experts, various attributes of the attacks.

Then, we extend the MulVAL attack graph generation and analysis framework to incorporate cyberattacks on ML production systems.
This extension considers both traditional attack techniques as well as adversarial ML attacks.
We conclude by demonstrating practical uses of the proposed extension, which will ultimately provide security practitioners with a tactical and practical tool for evaluating the impact and quantifying the risk of a cyberattack targeting ML systems.

The  main contributions of this research can be summarized as follows:
\begin{enumerate}
    \item An analysis of the main differences between a ML research pipeline and production pipeline.
    \item A methodological threat analysis of ML production system.
    \item An expert-based method for assigning a severity score for AML attacks.
    Using this method, we can highlight the attributes of an attack that contribute to the severity score as well as use the severity score in a risk assessment process. 
    \item An extension to the MulVAL attack graph generation and analysis framework that provides the ability to consider cyberattacks on ML production systems.
\end{enumerate}

\section{ML Production Systems}
\begin{figure*}[h]
\centering
\includegraphics[width=1.0\textwidth]{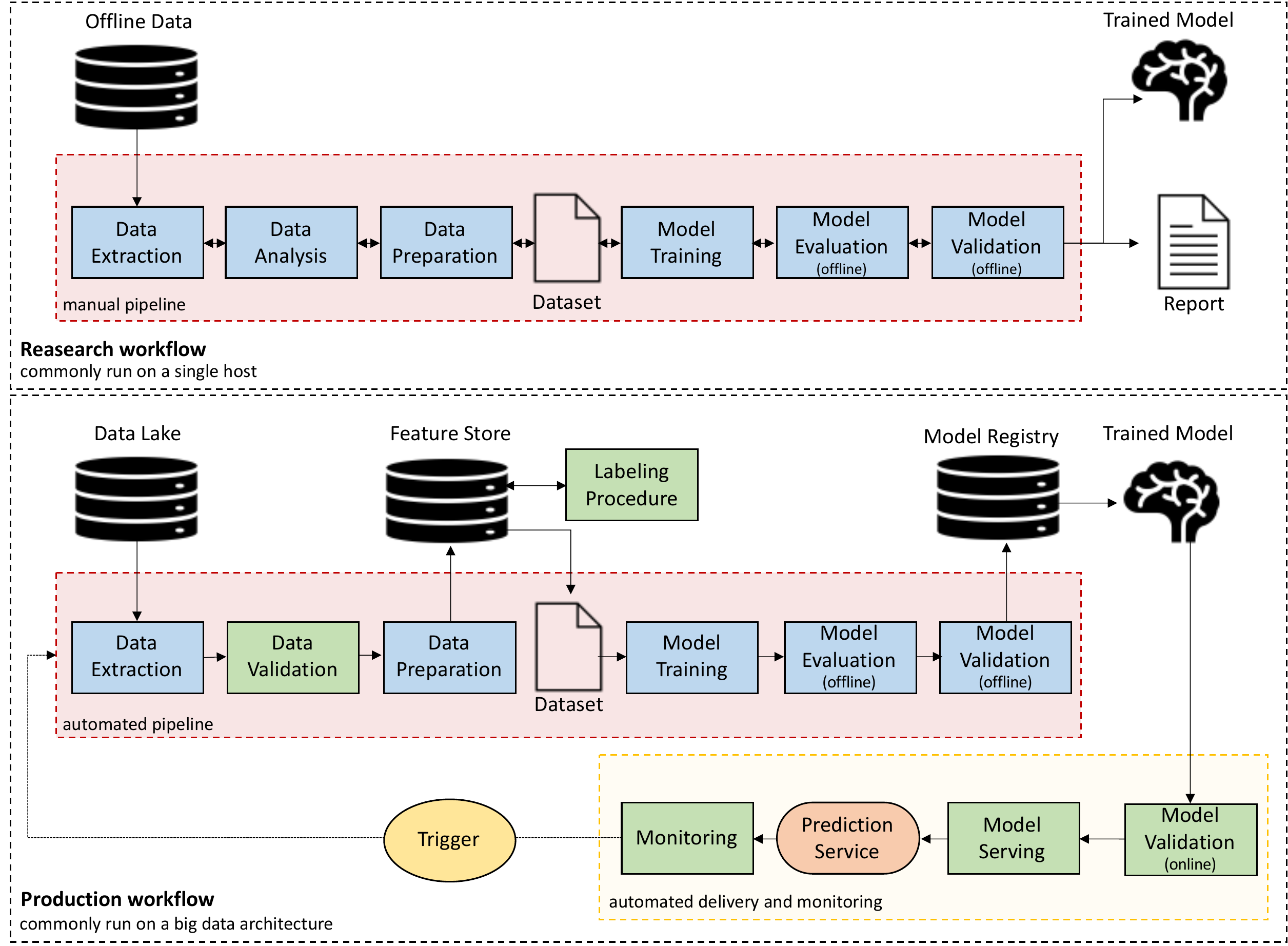}  \caption{Typical ML research and production pipelines.}
 \label{fig:mlops-ml-workflow}
\end{figure*}

ML systems are used in various use cases, each of which may have different deployment requirements and constraints.
In general, ML pipelines can be classified into two types: research pipeline and production pipeline (see Figure~\ref{fig:mlops-ml-workflow}). 
Understanding the characteristics of these two different pipelines is crucial for evaluating the risk of ML systems.
In this section, we discuss the main differences between a typical research pipeline and a typical production pipeline.
Our analysis is based on two reference environments: \textit{Michelangelo} -- Uber's ML platform,\footnote{\url{https://eng.uber.com/michelangelo-machine-learning-platform/}} and TensorFlow Extended -- Google's ML production platform~\cite{baylor2017tfx}.
The main characteristics and components of research and production pipelines are summarized in Table~\ref{tab:r_p_cmp}.

\begin{table*}[h!]
\scriptsize
\centering
\begin{tabular}{@{}l"c|c|c"c|c|c|c|c|c|c|c"b{0.003\textwidth}|b{0.003\textwidth}|b{0.003\textwidth}|b{0.003\textwidth}|b{0.003\textwidth}|@{}}
\Xhline{3\arrayrulewidth}
& Data & Deployment & Orchestration & \multicolumn{8}{c"}{Pipeline} & \multicolumn{5}{c|}{Delivery} \\ \cline{5-17} 
& & & &\rotatebox{90}{Data Extraction} & \rotatebox{90}{Data Analysis} & \rotatebox{90}{Data Validation} & \rotatebox{90}{Data Preparation}& \rotatebox{90}{Model Training} & \rotatebox{90}{Model Offline Evaluation} & \rotatebox{90}{Model Offline Validation} & \rotatebox{90}{Model Online Validation} & \rotatebox{90}{Continuous Model Retraining} & \rotatebox{90}{Performance Monitoring} & \rotatebox{90}{Feature Store} & \rotatebox{90}{Model Registry} & \rotatebox{90}{Model Serving} \\ \Xhline{3\arrayrulewidth}
\textbf{Research}    & Offline & Single Host & Manual & $\bullet$ & $\bullet$ & $\circ$  & $\bullet$  & $\bullet$ & $\bullet$ & $\bullet$ & $\circ$ & $\circ$ & $\circ$ & $\circ$ & $\circ$ & $\circ$ \\
\hline
\textbf{Production} & Online  & Cluster & Automated & $\bullet$ & $\circ$ & $\bullet$ & $\bullet$ & $\bullet$ & $\bullet$ & $\bullet$ & $\bullet$ & $\bullet$ & $\bullet$ & $\bullet$ & $\bullet$ & $\bullet$ \\ \Xhline{3\arrayrulewidth}
\end{tabular}%
\caption{A comparison of the main characteristics and components of machine learning research and production workflows.}
\label{tab:r_p_cmp}
\end{table*}

\subsection{The Characteristics of a Research Pipeline}
In a research pipeline, the main objectives are to explore novel ML applications that can benefit business goals and quickly develop proofs for that concepts. 
Therefore, in a research pipeline, the ML researcher (or data scientist) mainly focuses on extracting and analyzing data, defining new features, and testing (evaluating) different ML algorithms. 
The interactive nature of these tasks, coupled with the requirement for testing new concepts and algorithms quickly, results in the following characteristics of a typical ML research pipeline:

\begin{enumerate}[label=(\roman*)]

\item \textbf{Data.} In most cases, the data scientist works with static, offline datasets.
This way, the data scientist can explore the data and test new features quickly, using simple scripts, without the overhead of writing production-level code that connects to the enterprise's big data architecture components.

\item \textbf{Deployment.} The entire process (including data extraction, data analysis, data preparation, model training, and model evaluation) is implemented on a single host.

\item \textbf{Delivery.} 
The main outputs are a trained model and a report describing the application and evaluation results. 
The data scientist provides these outputs to the engineering team who serves the model as a prediction service in the production environment; i.e., the research pipeline does not include components required for using a predication model as a service.
Retraining the model with new data and actively monitoring model performance are tasks that are rarely considered in a research setup.
    
\item \textbf{Orchestration.} Every component (including data interpretation, data preparation, model training, and validation) is executed manually using dedicated scripts. 
In addition, the transition from one step to another is also performed manually. 

\end{enumerate}    

\subsection{The Characteristics of a Production Pipeline}

When ML models need to be deployed in production, engineering requirements (such as security, availability, throughput, scalability, compliance with privacy and bias regulations, etc.) must also be considered. 
Furthermore, when working with online real-world data, data trends often change over time. 
This behavior, known as concept drift, must be considered to ensure high performance over time.
Therefore, when ML needs to be deployed in production, the actual pipeline becomes much more complex:

\begin{enumerate}[label=(\roman*)]
\item \textbf{Data.} Large-scale live data must be utilized efficiently. 
Therefore, an ML production pipeline must be integrated into the enterprise's big data architectures.

\item \textbf{Deployment.} Data volumes can be tremendous. 
Therefore, a production pipeline is commonly deployed on big data architecture (i.e., a cluster of computers).

\item \textbf{Delivery.}
When working with online real-world data, data trends often change over time. 
Therefore, model performance must be actively monitored to detect performance degradation. 
In addition, to mitigate performance degradation, the ML models must be retrained frequently using fresh data. 
This way, the ML model can learn emerging patterns and trends over time.
However, before retraining a model, the training data must be validated to detect changes in the statistical properties of the data and identify data anomalies, which can cause the model to learn incorrect concepts. 
Furthermore, the retrained model must be re-evaluated (and validated) to ensure high performance.
A production pipeline also includes components to store and manage models and features (a model repository and feature store).

\item \textbf{Orchestration.} Since in a production pipeline models frequently change, model management must be automated and cannot be handled manually by a data scientist. 
This includes the execution of each step in the pipeline, as well as the transitions between steps.
\end{enumerate}

\section{The Security of ML Systems \label{sec:security_assessment}}
\begin{figure*}[!h]
\centering
\includegraphics[width=1\textwidth]{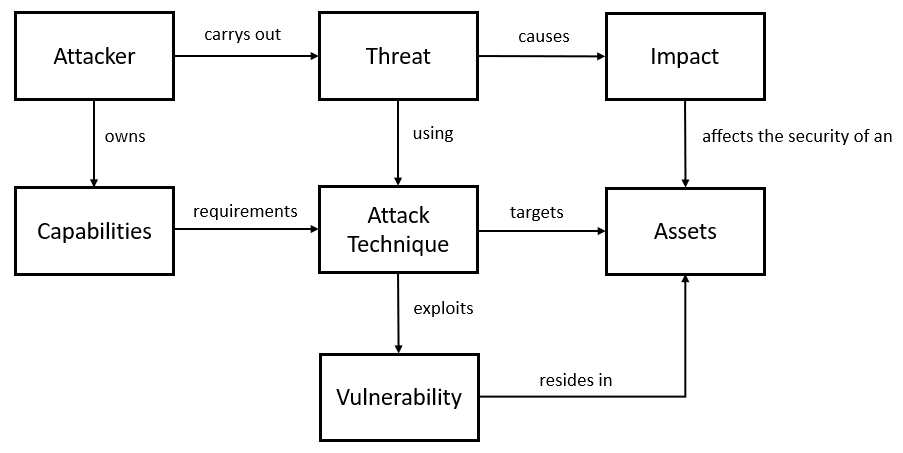}  
\caption{Threat analysis ontology.}
\label{fig:threat_ontology}
\end{figure*}

In this section, we analyze the security of ML production systems.
Our analysis is conducted according to a threat analysis ontology presented in Figure~\ref{fig:threat_ontology} which is based on the NIST ontology for evaluating enterprise security risk.

The ontology includes the following entities:
\begin{itemize}
    \item \textbf{Asset.} The main data, components, processes and services that are part of a ML production pipeline that should be protected. 
    
    \item \textbf{Vulnerability.} A characteristic of an asset or a technology that makes them prone to an attack.
    In the context to AML attacks, vulnerability refers to the inherent ability to systematically manipulate the input to the ML model.
    
    \item \textbf{Attacker.} An individual, group, or state responsible for an event or incident that impacts, or has the potential to impact, the security or safety of the system.
    Threat actors may have different capabilities and resources, and may perform different attacks.
    
    \item  \textbf{Capability.}
    Refers to the capabilities that are available to the attacker. 
    Within the context of AML attackers, we distinguish between access capabilities and knowledge capabiltiles.
    Knowledge capability access capability.
    
    \item \textbf{Impact.} By exploiting vulnerabilities, threats are able to cause a security impact on the vulnerable assets, e.g., violate the confidentiality of the data used for training a model or tempering with the ML model integrity. 
    Security impact may lead to operational impact. 
    
    \item \textbf{Threat.} Potential violation of a security property such as integrity, confidentiality, or availability.
    Threat, exploits some type of vulnerability, in order to conduct an attack. 
    
    \item \textbf{Attack Technique.} An act or a method that violates the security policy of a system.
\end{itemize}

According to the ontology, we begin by enumerating the assets of a typical ML production system; identifying these assets is a crucial step in threat modeling.
Then, we describe the threat model, which outlines the potential adversaries, their capabilities, and their main goals.
Next, we describe the various threats to ML systems.
Finally, we enumerate the attacks that can realize these threats.

\subsection{Target Assets \label{subsec:target_assets}}
A typical ML production system includes the following main components.

\noindent \textbf{[A1] Data Lake.} A big data architecture that is responsible for storing and processing large volumes of data.

\noindent \textbf{[A2] Data Extraction.} This component is responsible for selecting and integrating data from the various data sources that exist in the data lake.  

\noindent \textbf{[A3] Data Validation.} This component receives fresh data extracted from the data lake and is responsible for validating the input data to prevent the model from learning incorrect concepts.

\noindent \textbf{[A4] Data Preparation.} This component receives validated data and is responsible for preparing the data for the ML task. 
The preparation includes data cleansing, data transformations, and feature engineering.
In addition, this component is responsible for splitting the data into training, validation, and test sets. 

\noindent \textbf{[A5] Feature Store.} This component is a data warehouse responsible for storing and logging features for ML tasks. 
The main benefits of using this component are the ability to reuse available feature sets in different ML tasks; serve up-to-date feature values; and avoid training-serving skew by using the feature store as the data source for experimentation, continuous training, and online serving.

\noindent \textbf{[A6] Data Labeling.} This component (or process) is responsible for tagging data samples.
The process can be manual but is usually performed or assisted by software. 
The input for this process is commonly raw data, and predefined labeling logic, and the output consists of the labels of each data sample.

\noindent \textbf{[A7] Feature Selection.} This component is responsible for selecting the subset of features that will be used for training the model.

\noindent \textbf{[A8] Model Training.} This component is responsible for training ML algorithms with prepared data. 
The output of this step is a trained model.

\noindent \textbf{[A9] Model.} The trained model file, which is produced by feeding training data to the learning algorithm.

\noindent \textbf{[A10] Model Repository.} This component is a database responsible for storing models, their performance, versioning, and other configuration information (e.g., features used, hyperparameter values).

\noindent \textbf{[A11] Model Evaluation (offline).} This component is responsible for evaluating, using various measurements, the performance of a trained model on an unseen offline dataset (known as a test set).

\noindent \textbf{[A12] Model Validation (offline).} This component is responsible for validating that the new model is adequate for deployment (i.e., is better than a certain baseline). 

\noindent \textbf{[A13] Model Validation (online).} This component (also known as A/B testing) is responsible for evaluating, the performance of a trained model, on a small portion of unseen online data. 
The performance of the existing and new models are then compared, and the superior model is selected for serving.

\noindent \textbf{[A14] Model Serving.} This component is responsible for deploying the model to provide predictions (e.g., using a REST API).

\noindent \textbf{[A15] Model Performance Monitoring.} This component is responsible for actively monitoring model performance to detect performance degradation. 

\subsection{Threat Model \label{sec:threat_model}}
In this section, we define the threat model, which is based on the attacker's access capabilities and the knowledge regarding the target system.

\subsubsection{Attacker's access capabilities.  \label{subsubsec:attacker_access_capabilities}} 
\begin{figure*}[!h]
\centering
\includegraphics[width=1.0\textwidth]{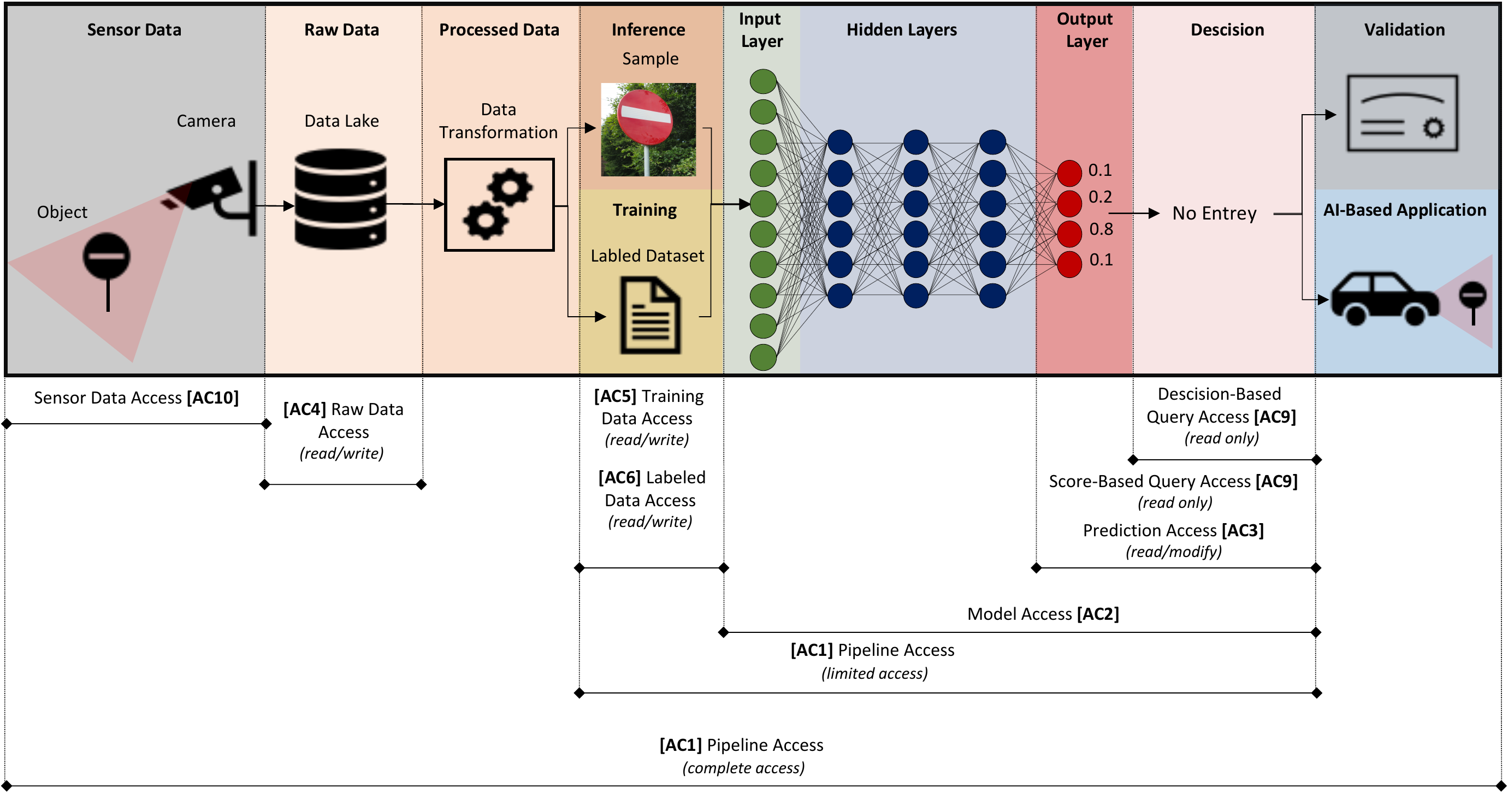}  
\caption{An illustration of the various access capabilities available to the adversary for different threat models.}
 \label{fig:threatmodel}
\end{figure*}

The set of capabilities accessible to the attacker regarding the target system (see Figure~\ref{fig:threatmodel}).

\noindent \textbf{[AC1] Pipeline Access.} In this threat model, we assume an adversary with access to all of the components included in the target pipeline, including the training data, algorithm, hyperparameters, model, and classification.

\noindent \textbf{[AC2] Model Access.} In this threat model, we assume an adversary that has access to the exact model used in the target pipeline.

\noindent \textbf{[AC3] Prediction Access.} In this threat model, we assume an adversary that has access to the outputs (predictions) of the model used in the target pipeline.

\noindent \textbf{[AC4] Raw Data Access.} In this threat model, we assume an adversary with access to the raw data (this data will be used for training the model after applying some data transformations).

\noindent \textbf{[AC5] Training Data Access.} In this threat model, we assume an adversary with access to the dataset used for training the target model.

\noindent \textbf{[AC6] Labeled Data Access.} In this threat model, we assume an adversary with access to the labels of the dataset used for training the target model.

\noindent \textbf{[AC7] Validation Data Access.} In this threat model, we assume an adversary with access to the dataset used for the external validation of the target model.

\noindent \textbf{[AC8] Surrogate Data Access.} In this threat model, we assume an adversary that has access to a reference dataset with similar characteristics and data distribution to the dataset used to train the model.

\noindent \textbf{[AC9] Score-Based Query Access.} In this threat model, we assume an adversary with the ability to query the trained model. 
We further assume that given a query, the attacker has access to the probability vector (i.e., score), which describes the confidence for each class. 
However, we do not assume the adversary has access to the training data.

\noindent \textbf{[AC10] Decision-Based Query Access.} In this threat model, we assume an adversary with the ability to query the trained model. 
We further assume that given a query, the attacker has access only to the decision. However, we do not assume the adversary has access to the probability vector or training data.

\noindent \textbf{[AC11] Sensor Data Access.} In this threat model, we assume an adversary with the ability to manipulate the data captured/measured by the sensors (e.g., an adversary that can manipulate an object that resides within the viewpoint of a camera).

\subsubsection{Attacker's knowledge. \label{subsubsec:attacker_knowledge}} 
Information regarding the target system that is available to the attacker and can be used to more effectively generate the attacks.

\noindent \textbf{[AK1] Perfect Knowledge.} In this threat model, we assume an adversary with complete knowledge on the target pipeline, including the training data, algorithm, hyperparameters, model, and classification.

\noindent \textbf{[AK2] Model Knowledge.} In this threat model, we assume an adversary that knows the exact model used in the target pipeline (e.g., the target pipeline uses a public model).

\noindent \textbf{[AK3] Hyperparameter Knowledge.} In this threat model, we assume an adversary that knows the exact algorithm and hyperparameters used to train the algorithm.
For example, for artificial neural networks, the hyperparameters include the network architecture, number of epochs used to train the model, selected learning rate, etc. 

\noindent \textbf{[AK4] Algorithm Knowledge.} In this threat model, we assume an adversary that knows the algorithm used to train the model, but does not know the exact hyperparameters.

\noindent \textbf{[AK5] Training Data Knowledge.} In this threat model, we assume an adversary that knows all/part of the training data used for training the model (including the exact feature transformations applied on the raw data data).

\noindent \textbf{[AK6] Raw Data Knowledge.} In this threat model, we assume an adversary that knows the exact data used for training the model (e.g., when a target model is being trained on a public dataset) but is not aware of the exact feature transformations applied to the raw data.

\noindent \textbf{[AK8] Data Property Knowledge.} In this threat model, we assume an adversary that knows the statistical properties of the data used to train the model.

\noindent \textbf{[AK9] Task Knowledge.} In this threat model, we assume an adversary that has general knowledge of the ML task, including the general type of inputs and outputs (e.g., the attacker knows that the pipeline is used to extract points of interest from the user's historical geolocations).

\subsubsection{Attacker's goal (or impact).\label{subsubsec:attacker_goal}}
Indicates what type of security impact (violation) the attacker wants to achieve.

\noindent \textbf{[AG1] Tampering.} This threat category is associated with malicious activities that would compromise the \textit{integrity} of the ML system.

\noindent \textbf{[AG2] Denial of Service.} This threat category is associated with malicious activities that would compromise the \textit{availability} of the ML system.
This goal can be achieved by causing the model to make a significant number of wrong predictions, and therefore, the model cannot be used, or sending specially crafted examples that results in a long prediction time by the model and/or high latency. 

\noindent \textbf{[AG3] Information Disclosure.} This threat category is associated with malicious activities that would compromise the \textit{privacy.} of the ML system.

\subsection{Threat Categories}
In this section, we enumerate the main threat categories to ML production systems.

\noindent \textbf{[T1] Evading an ML System.} An adversary causing the ML system to provide incorrect outputs for a specific input, for instance, causing a malware detection classifier to classify a malicious file as benign.

\noindent \textbf{[T2] ML Model Corruption.} An adversary causing the ML system to deploy a corrupted model, for instance, causing the ML system to deploy a corrupted malware detection classifier that includes a backdoor that will classify specific malicious files as benign.

\noindent \textbf{[T3] Membership Inference.} An adversary causing the ML system to leak information regarding the existence of a given input sample in the training set.  

\noindent \textbf{[T4] Data Property Inference.} An adversary causing the ML system to leak information in such a way that general properties of the training dataset (e.g., feature distribution, input types, etc.) are exposed.

\noindent \textbf{[T5] Data Reconstruction (theft).} An adversary causing the ML system to leak information in such a way that some of the data samples used to train the target model are exposed. 

\noindent \textbf{[T6] Model Extraction.} An adversary causing the ML system to leak information in such a way that the model used in the target pipeline is exposed.

\noindent \textbf{[T7] Wrong Prediction Flooding.} An adversary causing the ML system to make many wrong predictions rendering the ML model useless (e.g., a credit risk scoring model that provides wrong predictions or an IDS that raise many false alarms).

\noindent \textbf{[T8] Model DoS.} An adversary that can send specifically crafted examples that results in a long processing time by the ML system and eventually denies the service by other users or services.

\subsection{Attack Techniques \label{subsec:attack_tech}}
In this section, we review various attack techniques against ML systems. 
For each attack technique, we provide a brief description of the attack method, identify the assets targeted by the adversary, and mention the relevant threat model. 
A table summarizing the reviewed attacks is presented in Table~\ref{tab:attack_techniques}.

\setlength\tabcolsep{4.2pt}
\begin{table*}[h!]
\tiny
\centering
\hspace*{-1.4cm}
\begin{tabular}{@{}
|m{0.1\textwidth}"
m{0.28\textwidth}|
c"
b{0.005\textwidth}
b{0.005\textwidth}
b{0.005\textwidth}
b{0.005\textwidth}
b{0.005\textwidth}
b{0.005\textwidth}
b{0.005\textwidth}
b{0.005\textwidth}
b{0.005\textwidth}
b{0.005\textwidth}"
b{0.005\textwidth}
b{0.005\textwidth}
b{0.005\textwidth}
b{0.005\textwidth}
b{0.005\textwidth}
b{0.005\textwidth}
b{0.005\textwidth}
b{0.005\textwidth}"
b{0.005\textwidth}
b{0.005\textwidth}
b{0.005\textwidth}"
m{0.012\textwidth}|
@{}}
\Xhline{3\arrayrulewidth}

\makecell{Threat \\ Category} &
Attack Technique &
Target Assets &
\multicolumn{18}{c"}{\underline{Threat Model}} & 
\multicolumn{3}{c"}{\makecell{Attack \\ Impact}} &
S  \\

& & & 
\multicolumn{10}{c}{Access Capabilities} &
\multicolumn{8}{c"}{System Knowledge} & & & & \\ \cline{4-24}

& & & 
\rotatebox{90}{Pipeline Access $[AC1]$} & 
\rotatebox{90}{Model Access $[AC2]$} & 
\rotatebox{90}{Prediction Access $[AC3]$} & 
\rotatebox{90}{Raw Data Access $[AC4]$} &
\rotatebox{90}{Training Data Access $[AC5]$} &
\rotatebox{90}{Labeled Data Access $[AC6]$} &
\rotatebox{90}{Validation Data Access $[AC7]$} &
\rotatebox{90}{Surrogate Data Access $[AC8]$} &
\rotatebox{90}{Score-Based Query Access $[AC9]$} &
\rotatebox{90}{Decision-Based Query Access $[AC10]$} &

\rotatebox{90}{Perfect Knowledge $[AK1]$} &
\rotatebox{90}{Model Knowledge $[AK2]$} &
\rotatebox{90}{Hyperparameter Knowledge $[AK3]$} &
\rotatebox{90}{Algorithm Knowledge $[AK4]$} &
\rotatebox{90}{Training Data Knowledge $[AK5]$} &
\rotatebox{90}{Raw Data Knowledge $[AK6]$} &
\rotatebox{90}{Data Property Knowledge $[AK7]$} &
\rotatebox{90}{Task Knowledge $[AK8]$} &

\rotatebox{90}{Tampering $[AG1]$} &
\rotatebox{90}{Denial of Service $[AG2]$} &
\rotatebox{90}{Information Disclosure $[AG3]$} & \\ \Xhline{3\arrayrulewidth}

% \textit{[T1]} Evading an ML system
\multirow{9}{*}{\makecell{\textbf{Evading an ML} \\ \textbf{system \textit{[T1]}}}}  &
Gradient-based, white-box, evasion attacks (such as C\&W \cite{carlini2017adversarial}) FGSM \cite{goodfellow2014explaining}, and BIM \cite{kurakin2016adversarial}) &
$[A9]$ Model&
$\circ$ & $\circ$ & $\circ$  & $\circ$  & $\circ$ & $\circ$ & $\circ$ & $\circ$ & $\circ$ & $\circ$ &
$\bullet$ & $\bullet$ & $\bullet$ & $\bullet$ & $\bullet$ & $\bullet$ & $\bullet$ & $\bullet$ &
$\bullet$ & $\circ$  & $\circ$ & \cellcolor{Low} 6.9 \\ \cline{2-25}

&
Boundary-based, black-box (score-based), evasion attacks (such as ZOO \cite{chen2017zoo})&
$[A9]$ Model&
$\circ$ & $\circ$ & $\circ$  & $\circ$  & $\circ$ & $\circ$ & $\circ$ & $\circ$ & $\bullet$ & $\circ$ &
$\circ$ & $\circ$ & $\circ$ & $\circ$ & $\circ$ & $\circ$ & $\circ$ & $\bullet$ &
$\bullet$ & $\circ$  & $\circ$ & \cellcolor{VeryHigh} 7.6 \\ \cline{2-25}

&
Boundary-based, black-box (decision-based), evasion attacks (such as HopSkipJump \cite{chen2020hopskipjumpattack})&
$[A9]$ Model&
$\circ$ & $\circ$ & $\circ$  & $\circ$  & $\circ$ & $\circ$ & $\circ$ & $\circ$ & $\circ$ & $\bullet$ &
$\circ$ & $\circ$ & $\circ$ & $\circ$ & $\circ$ & $\circ$ & $\circ$ & $\bullet$ &
$\bullet$ & $\circ$  & $\circ$ & \cellcolor{VeryHigh} 7.9  \\ \cline{2-25}

&
Transferability-based, black-box (decision-based), evasion attacks that utilize reference data (such as Jacobian Data Augmentation \cite{papernot2016transferability})&
$[A9]$ Model&
$\circ$ & $\circ$ & $\circ$  & $\circ$  & $\circ$ & $\circ$ & $\circ$ & $\odot$ & $\circ$ & $\bullet$ &
$\circ$ & $\circ$ & $\circ$ & $\circ$ & $\circ$ & $\circ$ & $\bullet$ & $\bullet$ &
$\bullet$ & $\circ$  & $\circ$  & \cellcolor{High} 7.1 \\ \cline{2-25}

&
Transferability-based, black-box (decision-based), evasion attacks that utilize training data (such as \cite{papernot2016transferability})&
$[A9]$ Model&
$\circ$ & $\circ$ & $\circ$  & $\circ$  & $\circ$ & $\circ$ & $\circ$ &  $\circ$ & $\circ$ & $\bullet$ &
$\circ$ & $\circ$ & $\circ$ & $\circ$ & $\bullet$ & $\bullet$ & $\bullet$ & $\bullet$ &
$\bullet$ & $\circ$  & $\circ$  & \cellcolor{Medium} 7.6 \\ \cline{2-25}

&
Gradient-based, targeted, white-box, poisoning attack (such as  \cite{kravchik2021poisoning,mei2015using,jagielski2018manipulating})&
$[A9]$ Model&
$\circ$ & $\circ$ & $\circ$  & $\odot$  & $\odot$ & $\odot$ & $\circ$ & $\circ$ & $\circ$ &$\odot$ &
$\bullet$ & $\bullet$ & $\bullet$ & $\bullet$ & $\bullet$ & $\bullet$ & $\bullet$ & $\bullet$ &
$\bullet$ & $\circ$  & $\circ$  & \cellcolor{VeryLow} 5.6\\ \cline{2-25}

&
Transferability-based, targeted, black-box, poisoning attack (error-specific or error generic such as \cite{biggio2012poisoning,munoz2017towards,jagielski2018manipulating}))&
$[A9]$ Model&
$\circ$ & $\circ$ & $\circ$  & $\odot$  & $\odot$ & $\odot$ & $\circ$ &  $\bullet$ & $\circ$ &$\odot$ &
$\circ$ & $\circ$ & $\circ$ & $\circ$ & $\bullet$ & $\circ$ & $\bullet$ & $\bullet$ &
$\bullet$ & $\circ$  & $\circ$  & \cellcolor{Low} 6.3 \\ \cline{2-25}

&
Gradient-based, targeted, white-box, poisoning attack against feature selection (such as \cite{xiao2015feature})&
$[A7]$ Feature selection&
$\circ$ & $\circ$ & $\circ$  & $\odot$  & $\odot$ & $\odot$ & $\circ$ & $\circ$ & $\circ$ &$\odot$ &
$\bullet$ & $\bullet$ & $\bullet$ & $\bullet$ & $\bullet$ & $\bullet$ & $\bullet$ & $\bullet$ &
$\bullet$ & $\circ$  & $\circ$  & \cellcolor{VeryLow} 5.6  \\ \cline{2-25}

&
Transferability-based, targeted, black-box, poisoning attack against feature selection (such as \cite{xiao2015feature})&
$[A7]$ Feature selection&
$\circ$ & $\circ$ & $\circ$  & $\odot$  & $\odot$ & $\odot$ & $\circ$ &  $\bullet$ & $\circ$ &$\odot$ &
$\circ$ & $\circ$ & $\circ$ & $\circ$ & $\bullet$ & $\circ$ & $\bullet$ & $\bullet$ &
$\bullet$ & $\circ$  & $\circ$  & \cellcolor{Low} 6.3  \\ \Xhline{3\arrayrulewidth}

% \textit{[T2]} ML model corruption

\multirow{4}{*}{\makecell{\textbf{ML model} \\ \textbf{corruption \textit{[T2]}}}} &
Gradient-based, indiscriminate, white-box poisoning attack (such as \cite{kravchik2021poisoning,mei2015using,jagielski2018manipulating})&
$[A9]$ Model&
$\circ$ & $\circ$ & $\circ$  & $\odot$  & $\odot$ & $\odot$ & $\circ$ & $\circ$ & $\circ$ &$\odot$ &
$\bullet$ & $\bullet$ & $\bullet$ & $\bullet$ & $\bullet$ & $\bullet$ & $\bullet$ & $\bullet$ &
$\bullet$ & $\bullet$  & $\circ$  & \cellcolor{VeryLow} 5.6 \\ \cline{2-25}

&
Transferability-based, indiscriminate, black-box poisoning attack (such as \cite{biggio2012poisoning,munoz2017towards,jagielski2018manipulating})&
$[A9]$ Model&
$\circ$ & $\circ$ & $\circ$  & $\odot$  & $\odot$ & $\odot$ & $\circ$ &  $\bullet$ & $\circ$ &$\odot$ &
$\circ$ & $\circ$ & $\circ$ & $\circ$ & $\bullet$ & $\circ$ & $\bullet$ & $\bullet$ &
$\bullet$ & $\bullet$  & $\circ$  & \cellcolor{Low} 6.8 \\ \cline{2-25}

&
Gradient-based, indiscriminate, white-box poisoning attack against feature selection (such as \cite{xiao2015feature})&
$[A7]$ Feature selection&
$\circ$ & $\circ$ & $\circ$  & $\odot$  & $\odot$ & $\odot$ & $\circ$ & $\circ$ & $\circ$ &$\odot$ &
$\bullet$ & $\bullet$ & $\bullet$ & $\bullet$ & $\bullet$ & $\bullet$ & $\bullet$ & $\bullet$ &
$\bullet$ & $\bullet$  & $\circ$   & \cellcolor{VeryLow} 5.6 \\ \cline{2-25}

&
Transferability-based, indiscriminate, black-box poisoning attack against feature selection (such as \cite{xiao2015feature})&
$[A7]$ Feature selection&
$\circ$ & $\circ$ & $\circ$  & $\odot$  & $\odot$ & $\odot$ & $\circ$ &  $\bullet$ & $\circ$ &$\odot$ &
$\circ$ & $\circ$ & $\circ$ & $\circ$ & $\bullet$ & $\circ$ & $\bullet$ & $\bullet$ &
$\bullet$ & $\bullet$  & $\circ$  & \cellcolor{Low} 6.8 \\ \Xhline{3\arrayrulewidth}

%\textit{[T3]} Membership inference}

\multirow{3}{*}{\makecell{\textbf{Membership} \\ \textbf{inference \textit{[T3]}}}}&
Shadow training-based, black-box membership inference attacks \cite{shokri2017membership}&
\makecell{ $[A1]$ Data lake  \\ $[A5]$ Feature store }&
$\circ$ & $\circ$ & $\circ$  & $\odot$  & $\odot$ & $\circ$ & $\circ$ & $\circ$ & $\bullet$ & $\bullet$ &
$\circ$ & $\circ$ & $\circ$ & $\circ$ & $\odot$ & $\circ$ & $\circ$ & $\bullet$ &
$\circ$ & $\circ$  & $\bullet$  & \cellcolor{VeryHigh} 7.7 \\ \cline{2-25}

&
Gradient-based, white-box membership inference attacks (such as \cite{nasr2019comprehensive})&
\makecell{ $[A1]$ Data lake  \\ $[A5]$ Feature store }&
$\circ$ & $\bullet$ & $\bullet$  & $\odot$  & $\odot$ & $\circ$ & $\circ$ & $\circ$ & $\bullet$ & $\circ$ &
$\circ$ & $\bullet$ & $\bullet$ & $\bullet$ & $\odot$ & $\circ$ & $\circ$ & $\bullet$ &
$\circ$ & $\circ$  & $\bullet$  & \cellcolor{Low} 6.8 \\ \cline{2-25}

&
Membership inference attacks exploiting attacker's access &
\makecell{ $[A1]$ Data lake  \\ $[A5]$ Feature store }&
$\circ$ & $\circ$ & $\circ$  & $\bullet$  & $\bullet$ & $\bullet$ & $\circ$ & $\circ$ & $\circ$ & $\circ$ &
$\circ$ & $\circ$ & $\circ$ & $\circ$ & $\circ$ & $\circ$ & $\circ$ & $\circ$ &
$\circ$ & $\circ$  & $\bullet$  & \cellcolor{VeryHigh} 7.0 \\ \Xhline{3\arrayrulewidth}

%\textit{[T4]} Data property inference} 
\multirow{2}{*}{\makecell{\textbf{Data property} \\ \textbf{inference \textit{[T4]}}}}&
Shadow training-based property inference attacks \cite{ateniese2015hacking,ganju2018property,song2020information}&
\makecell{ $[A1]$ Data lake  \\ $[A5]$ Feature store }&
$ \circ $ & $ \circ $ & $ \circ $  & $ \odot $  & $ \odot $ & $ \circ $ & $ \circ $ & $ \circ $ & $ \bullet $ & $ \bullet $ &
$ \circ $ & $ \circ $ & $ \circ $ & $ \circ $ & $ \odot $ & $ \circ $ & $ \circ $ & $ \odot $ &
$ \circ $ & $ \circ $  & $ \bullet $  & \cellcolor{VeryHigh} 7.7  \\ 
\cline{2-25}

&
Property inference attacks exploiting attacker's access &
\makecell{ $[A1]$ Data lake  \\ $[A5]$ Feature store }&
$\circ$ & $\circ$ & $\circ$  & $\bullet$  & $\bullet$ & $\bullet$ & $\circ$ & $\circ$ & $\circ$ & $\circ$ &
$\circ$ & $\circ$ & $\circ$ & $\circ$ & $\circ$ & $\circ$ & $\circ$ & $\circ$ &
$\circ$ & $\circ$  & $\bullet$  & \cellcolor{High} 7.0  \\ \Xhline{3\arrayrulewidth}

% \textbf{\textit{[T4]} Data property inference} &
% White-box property inference attacks \cite{}&
% 
% \textbf{I}&
% $ \circ $ & $ \bullet $ & $ \bullet $  & $ \odot $  & $ \odot $ & $ \circ $ & $ \circ $ & $ \odot $ & $ \bullet $ & $ \bullet $ &
% $ \odot $ & $ \bullet $ & $ \bullet $ & $ \bullet $ & $ \odot $ & $ \circ $ & $ \circ $ & $ \odot $ &
% $ \circ $ & $ \circ $  & $ \bullet $ \\ \hline

%\textit{[T5]} Data reconstruction (theft)

\multirow{4}{*}{\makecell{\textbf{Data} \\ \textbf{reconstruction} \\ \textbf{\textit{[T5]}}}} &
Maximum a posteriori-based, gray-box data reconstruction attacks \cite{fredrikson2014privacy,hidano2017model}&
\makecell{ $[A1]$ Data lake  \\ $[A5]$ Feature store }&
$\circ$ & $\odot$ & $\odot$  & $\odot$  & $\odot$ & $\circ$ & $\circ$ & $\circ$ & $\bullet$ & $\bullet$ &
$\circ$ & $\bullet$ & $\circ$ & $\circ$ & $\odot$ & $\odot$ & $\circ$ & $\odot$ &
$\circ$ & $\circ$  & $\bullet$  & \cellcolor{High} 7.0 \\ \cline{2-25}

&
Gradient optimization-based, gray-box data reconstruction attacks \cite{fredrikson2015model}&
\makecell{ $[A1]$ Data lake  \\ $[A5]$ Feature store }&
$\circ$ & $\odot$ & $\odot$  & $\odot$  & $\odot$ & $\circ$ & $\circ$ & $\circ$ & $\bullet$ & $\bullet$ &
$\circ$ & $\bullet$ & $\circ$ & $\circ$ & $\odot$ & $\odot$ & $\circ$ & $\odot$ &
$\circ$ & $\circ$  & $\bullet$ & \cellcolor{High} 7.0 \\ \cline{2-25}

&
Black-box data reconstruction attacks \cite{yang2019neural}&
\makecell{ $[A1]$ Data lake  \\ $[A5]$ Feature store }&
$ \circ $ & $ \circ $ & $ \circ $  & $ \odot $  & $ \odot $ & $ \circ $ & $ \circ $ & $ \circ $ & $ \bullet $ & $ \bullet $ &
$ \circ $ & $ \circ $ & $ \circ $ & $ \circ $ & $ \odot $ & $ \odot $ & $ \circ $ & $ \odot $ &
$ \circ $ & $ \circ $  & $ \bullet $ & \cellcolor{VeryHigh} 7.5 \\ \cline{2-25}

&
Data theft due to data breach &
\makecell{ $[A1]$ Data lake  \\ $[A5]$ Feature store }&
$\circ$ & $\circ$ & $\circ$  & $\bullet$  & $\bullet$ & $\bullet$ & $\circ$ & $\circ$ & $\circ$ & $\circ$ &
$\circ$ & $\circ$ & $\circ$ & $\circ$ & $\circ$ & $\circ$ & $\circ$ & $\circ$ &
$\circ$ & $\circ$  & $\bullet$ & \cellcolor{High} 7.0 \\ \Xhline{3\arrayrulewidth}

\multirow{2}{*}{\makecell{\textbf{Model} \\ \textbf{extraction \textit{[T6]}}}} &
Query-based, black-box model extraction attacks \cite{chandrasekaran2020exploring, juuti2019prada, oh2019towards, papernot2017practical}&
$[A9]$ Model&
$\circ$ & $\circ$ & $\circ$  & $\circ$  & $\circ$ & $\circ$ & $\circ$ & $\circ$ & $\bullet$ & $\bullet$ &
$\circ$ & $\circ$ & $\circ$ & $\circ$ & $\odot$ & $\odot$ & $\circ$ & $\odot$ &
$\circ$ & $\circ$  & $\bullet$ & \cellcolor{VeryHigh} 8.0\\ \cline{2-25}

&
Attacker's access-based model extraction attack &
$[A9]$ Model&
$\circ$ & $\bullet$ & $\circ$  & $\circ$  & $\circ$ & $\circ$ & $\circ$ & $\circ$ & $\circ$ & $\circ$ &
$\circ$ & $\circ$ & $\circ$ & $\circ$ & $\circ$ & $\circ$ & $\circ$ & $\circ$ &
$\circ$ & $\circ$  & $\bullet$  & \cellcolor{High} 7.0\\ \Xhline{3\arrayrulewidth}

\multicolumn{25}{l}{Access capabilities -  \textbf{$\circ$}: does not require such type of access
 , \textbf{$\odot$}: limited access of such type is required , \textbf{$\bullet$}: complete access of such type is required.} \\

\multicolumn{25}{l}{System knowledge -  \textbf{$\circ$}: does not require such type of knowledge,$\odot$: require partial knowledge \textbf{$\bullet$}: require full knowledge.} \\
\multicolumn{25}{l}{Attack Goal -  \textbf{$\circ$}: The attacker can achieve such type of goal, by executing this attack, \textbf{$\bullet$}: The attacker can't achieve such type of goal, by executing this attack.} \\

\end{tabular}%
\caption{Adversarial machine learning attack techniques.}
\label{tab:attack_techniques}
\end{table*}

\noindent\textbf{Evasion attacks \label{subsubsec:evasion_attacks}} 
In evasion attacks the adversary exploits the ML model \textit{[A9]} by generating a crafted input sample (an adversarial example) which is very similar to some other correctly classified input but is incorrectly classified by the ML model.
The adversary's main objective is to compromise the integrity \textit{[AG1]} of the ML model \textit{[A9]} by causing the ML system to provide incorrect outputs for a specific input \textit{[T1]}. 
Evasion attacks can be broadly classified based on the following criteria: attack technique, which characterizes the technique used by the attacker to craft the adversarial example (e.g., gradient-based, boundary-based, and transferability-based); the threat model, which characterizes the attacker’s access capabilities and attacker's knowledge of the ML-based system  (e.g., white-box, black-box); and the attack's specificity, which characterizes the goal of the attacker (e.g., targeted vs untargeted).  

\noindent \textbf{[AT1] Gradient-based white-box evasion attacks (targeted and untargeted).}
In these types of attacks, the adversary manipulates an input sample in such a way that the classification loss is maximized.
The specific manipulation is determined by calculating the gradients of the classification loss with respect to the input sample~\cite{goodfellow2014explaining,kurakin2016adversarial,carlini2017adversarial}.
For example, in the FGSM attack~\cite{goodfellow2014explaining}, the adversarial example is calculated by adding noise to the input sample, in the direction of the gradient of the classification loss with respect to the input sample.
Since these types of attacks are based on gradient computation, 
they consider a white-box adversary with perfect knowledge of the target pipeline \textit{[AK1]}. 
In addition, they further assume that the adversary has the ability to query the trained model \textit{[AC9][AC10]}; this requirement is necessary for executing the attack (i.e., sending the crafted adversarial example for classification).

\noindent \textbf{[AT2] Boundary-based black-box (score-based) evasion attacks (targeted and untargeted).}
Similar to \textit{[AT1]}, these types of attacks generate the adversarial example based on the gradients of the classification loss with respect to the input sample, however in contrast to \textit{[AT1]} that \textit{calculate} the gradients, these attacks \textit{estimate} the gradients.
For example, in the ZOO attack~\cite{chen2017zoo}, the adversary utilizes zeroth-order optimization methods to directly estimate the gradients of the target model.
Since these attacks are not based on gradient computation, they consider an adversary with general knowledge of the task \textit{[AK8]} and black-box score-based query access \textit{[AC9]}.

\noindent \textbf{[AT3] Boundary-based black-box (decision-based) evasion attacks (targeted and untargeted).}
Similar to \textit{[AT2]}, these types of attacks also \textit{estimate} the gradients.
However, in contrast to \textit{[AT2]} that estimate the gradient by using the classification score (i.e., vector of probabilities), these attacks estimate the gradients using the label alone (without using the vector of probabilities).
For example, in the HopSkipJump attack~\cite{chen2017zoo}, the adversary utilizes Monte Carlo methods to directly estimate the gradients of the target model.
As a result, these attacks consider an adversary with general knowledge of the task \textit{[AK8]} and black-box decision-based query access \textit{[AC10]}.

\noindent \textbf{[AT4] Transferability-based black-box (decision-based) evasion attacks that utilize reference data (targeted and untargeted).}
These types of attacks include the following three main phases:
First, the adversary creates a surrogate model by training a learning algorithm on a reference dataset with similar characteristics and distribution as the training set \textit{[AC8]}.
Second, the adversary generates adversarial examples by executing white-box gradient-based attacks on the surrogate model.
Third, the adversary uses the adversarial examples (generated against the surrogate model) to attack the target model.
These attacks exploit the \textit{transferability} property of ML models~\cite{papernot2016transferability,szegedy2013intriguing,goodfellow2014explaining}, i.e., adversarial examples that affect one model can often affect other models, even if they trained using different learning algorithms, hyperparameters, or training sets, as long as all models were trained to perform the same task (e.g., image classification).
For example in~\cite{papernot2016transferability}, the adversary utilizes Jacobian data augmentation to generate a surrogate dataset given an initial substitute training set of limited size.  
Since the gradient-based attacks are executed on a surrogate model, they can be executed by a black-box attacker with very limited query access \textit{[AC10]} as long as the adversary knows the general task \textit{[AK8]} and data properties [AK7] and has access to a reference dataset \textit{[AC8]}.

\noindent \textbf{[AT5] Transferability-based black-box (decision-based) evasion attacks (targeted and untargeted) that utilize training data.} 
Similar to \textit{[AT4]}, the adversary executes gradient-based attacks on a surrogate model (created by the attacker).
The only difference is that in these attacks, the adversary creates a surrogate model by training a learning algorithm on the exact dataset used to train the target model rather than on a reference dataset.
Therefore, as long as the adversary knows the training data used to train the model \textit{[AK5]\textit{[AK6]}\textit{[AK7]}} and has knowledge on the general task \textit{[AK8]}, these attacks can be executed by a black-box attacker with very limited query access \textit{[AC10]}.

\noindent\textbf{Poisoning attacks \label{subsubsec:poisoning_attacks}} 
In poisoning attacks, the adversary exploits the ML training procedure \textit{[A8]}, by injecting crafted data samples into the dataset used for training the learning algorithm \textit{[AG2]}.
The adversary main objective is either compromising the availability \textit{[AG2]} of the ML model \textit{[A9]} by causing the ML system to deploy a corrupted model \textit{[T2]}; or compromising the integrity \textit{[AG1]} of the ML model \textit{[A9]}, by causing the ML system to deploy a model, which provides wrong outputs on a specific input \textit{[T1]}.
Poisoning attacks can be generally classified based on the following criteria: attack technique, which characterizes the technique used by the attacker to craft the malicious input samples (i.e., gradient-based, if the attack is based on a gradient optimization method; and transferability-based, if the attack utilizes the transferability property of ML models); threat model, which characterizes the attacker’s access capabilities and the attacker's knowledge regarding the system (e.g., white-box and black-box), attack's specificity, which characterizes the goal of the attacker (i.e., targeted, if the attack aims to cause misclassification of a specific set of samples; or indiscriminate, if the attack aims to cause misclassification of any sample), and error specificity, which characterizes the type of error in multiclass problems (i.e., error-specific, if the attacker aims to have a sample misclassified as a specific class; or error-generic, if the attacker aims to have a sample misclassified as any of the classes other than the true class).

\noindent \textbf{[AT6] Gradient-based targeted white-box poisoning attacks (error-specific and error-generic).}
In these types of attacks, the adversary manipulates a small number of training samples in such a way that the classification loss of some other set of samples targeted by the attacker is maximized \textit{[T1],[AG1]}.
The specific manipulation is calculated by solving, using gradient-optimization techniques, a bilevel optimization problem where the outer optimization aims at manipulating the malicious input to maximize the loss function on a dataset that includes the samples targeted by the attacker, while the inner optimization corresponds to retraining the learning algorithm on a dataset that includes the malicious examples~\cite{kravchik2021poisoning,mei2015using,jagielski2018manipulating}.
Since these types of attacks are based on the adversary's ability to manipulate a small number of training samples, the adversary must have the ability to write to the dataset used to train the model \textit{[AC3],[AC4],[AC5]}.   
In addition, since these attacks are based on gradient computation, the adversary must possess perfect knowledge of the target pipeline (including the model architecture and parameters) \textit{[AK1]}. 
Furthermore, to execute the attack, the adversary must have the ability to query the trained model \textit{[AC10]}. 

\noindent \textbf{[AT7] Transferability-based targeted black-box poisoning attacks (error-specific and error-generic).}
Similar to \textit{[AT6]}, the adversary manipulates a small number of training samples in a such a way that the classification loss of some other set of samples targeted by the attacker is maximized \textit{[T1],[AG1]}.
Therefore, these types of attacks also require the adversary to have the ability to write into the dataset used to train the model \textit{[AC3],[AC4],[AC5]}.   
However, in contrast to \textit{[AT6]} which calculate the gradients using the target model (which requires perfect knowledge of the system), these attacks calculate the gradients using a surrogate model~\cite{biggio2012poisoning,munoz2017towards,jagielski2018manipulating} (created by training a learning algorithm on a reference dataset whose  characteristics and distribution are similar to the training set) \textit{[AC8]}.
As a result, they can be executed by a black-box attacker with very limited query access \textit{[AC10]}, as long as the adversary knows the general task \textit{[AK8]} and data properties [AK7], and has to access to a reference dataset \textit{[AC8]}.

\noindent \textbf{[AT8] Gradient-based targeted white-box poisoning attacks against feature selection (error-specific and error-generic).}
In these types of attacks, the adversary manipulates a small number of training samples in a way that leads the feature selection algorithm to select a specific subset of features targeted by the attacker \textit{[T1],[AG1]}.
The specific manipulation is calculated by solving, using gradient-optimization techniques, a bilevel optimization problem where the outer optimization aims at manipulating the malicious input to maximize the feature selection loss function on a dataset that includes the samples targeted by the attacker, while the inner optimization corresponds to retraining the feature selection algorithm on a dataset that includes the malicious examples~\cite{xiao2015feature}.
Since these types of attacks are based on the adversary's ability to manipulate a small number of training samples, the adversary must have the ability to write into the dataset used for training the model \textit{[AC3],[AC4],[AC5]}.   
In addition, since these attacks are based on gradient computation, the adversary must possess perfect knowledge of the target pipeline (including model architecture and parameters) \textit{[AK1]}. 
Furthermore, to execute the attack, the adversary must have the ability to query the trained model \textit{[AC10]}. 

\noindent \textbf{[AT9] Transferability-based targeted black-box poisoning attacks against feature selection (error-specific and error-generic).}
Similar to \textit{[AT6]}, the adversary manipulates a small number of training samples in a way that leads the feature selection algorithm to select a specific subset of features targeted by the attacker \textit{[T1],[AG1]}.
Therefore, these types of attacks also require the adversary to have the ability to write to the dataset used to train the model \textit{[AC3],[AC4],[AC5]}.   
However, in contrast to \textit{[AT6]} which calculate the gradients using the target model (which requires perfect knowledge of the system), these attacks calculate the gradients using a surrogate model \cite{xiao2015feature} \textit{[AC8]}.
As a result, they can be executed by a black-box attacker with very limited query access \textit{[AC10]}, as long as the adversary knows the general task \textit{[AK8]} and data properties [AK7], and has to access to reference dataset \textit{[AC8]}.

\noindent \textbf{[AT10] Gradient-based indiscriminate white-box poisoning attacks (error-specific and error-generic).}
In these types of attacks, the adversary manipulates a small number of training samples in a such a way that the classification loss of an untainted dataset is maximized \textit{[T1],[AG2]}.
The specific manipulation is calculated by solving, using gradient-optimization techniques, a bilevel optimization problem where the outer optimization aims at manipulating the malicious input to maximize the loss function on a dataset that includes the samples targeted by the attacker, while the inner optimization corresponds to retraining the learning algorithm on an untainted dataset (e.g., a validation dataset, which does not include malicious examples) \cite{kravchik2021poisoning,mei2015using,jagielski2018manipulating}.
Since these types of attacks are based on the adversary's ability to manipulate a small number of training samples, the adversary must have the ability to write into the dataset used to train the model \textit{[AC3],[AC4],[AC5]}.   
In addition, since these attacks are based on gradient computation, the adversary must possess perfect knowledge of the target pipeline (including the model architecture and parameters) \textit{[AK1]}. 
Furthermore, to execute the attack, the adversary must have the ability to query the trained model \textit{[AC10]}. 

\noindent \textbf{[AT11] Transferability-based indiscriminate black-box poisoning attacks (error-specific and error-generic).}
Similar to \textit{[AT10]}, the adversary manipulates a small number of training samples in a such a way that the classification loss of an untainted dataset is maximized \textit{[T1],[AG2]}.
Therefore, these types of attacks also require the adversary to have the ability to write into the dataset used to train the model \textit{[AC3],[AC4],[AC5]}.   
However, in contrast to \textit{[AT10]} which calculate the gradients using the target model (which requires perfect knowledge of the system), these attacks calculate the gradients using a surrogate model~\cite{biggio2012poisoning,munoz2017towards,jagielski2018manipulating} (created by training a learning algorithm on a reference dataset whose characteristics and distribution are similar to the training set) \textit{[AC8]}.
As a result, they can be executed by a black-box attacker with very limited query access \textit{[AC10]}, as long as the adversary knows the general task \textit{[AK8]} and data properties [AK7], and has to access to reference dataset \textit{[AC8]}.

\noindent \textbf{[AT12] Gradient-based indiscriminate white-box poisoning attacks against feature selection (error-specific and error-generic).}
In these types of attacks, the adversary manipulates a small number of training samples in a way that leads the feature selection algorithm to select a different subset of features \textit{[T1],[AG2]}.
The specific manipulation is calculated by solving, using gradient-optimization techniques, a bilevel optimization problem where the outer optimization aims at manipulating the malicious input to maximize the feature selection loss function on a dataset that includes the samples targeted by the attacker, while the inner optimization corresponds to retraining the feature selection algorithm on an untainted dataset~\cite{xiao2015feature}.
Since these types of attacks are based on the adversary's ability to manipulate a small number of training samples, the adversary must have the ability to write into the dataset used to train the model \textit{[AC3],[AC4],[AC5]}.   
In addition, since these attacks are based on gradient computation, the adversary must possess perfect knowledge of the target pipeline (including the model architecture and parameters) \textit{[AK1]}. 
Furthermore, to execute the attack, the adversary must have the ability to query the trained model \textit{[AC10]}. 

\noindent \textbf{[AT13] Transferability-based indiscriminate black-box poisoning attacks against feature selection (error-specific and error-generic).}
Similar to \textit{[AT6]}, the adversary manipulates a small number of training samples in a way that leads the feature selection algorithm to select a specific subset of features targeted by the attacker \textit{[T1]}.
Therefore, these types of attacks also require the adversary to have the ability to write into the dataset used to train the model \textit{[AC3],[AC4],[AC5]}.   
However, in contrast to \textit{[AT10]} which calculate the gradients using the target model (which requires perfect knowledge of the system), transferability-based poisoning attacks calculate the gradients using a surrogate model (created by running the feature selection algorithm on a reference dataset whose characteristics and distribution are similar to the training set) \textit{[AC8]}.
As a result, they can be executed by a black-box attacker with very limited query access \textit{[AC10]}, as long as the adversary knows the general task \textit{[AK8]} and data properties [AK7], and has to access to reference dataset \textit{[AC8]}.

\noindent\textbf{Inference attacks \label{subsubsec:inference_attacks}}
In inference attacks (such as membership inference \textit{[T3]}, data property inference \textit{[T4]}, data reconstruction \textit{[T5]}, and model reconstruction \textit{[T6]}), the adversary exploits the ML model \textit{[A9]} to expose private information and thereby compromise the privacy \textit{[AG3]} of the ML model \textit{[A9]} or the data used to train it \textit{[A1],[A5]}.

\noindent \textbf{[AT14] Shadow training-based black-box membership inference attacks.}
In these types of attacks, given an ML model \textit{[A9]} and a record, the adversary's goal is to determine whether the record was used as part of the model's training dataset \textit{[T3]} and thereby compromise the privacy of that record \textit{[AG3]}. These attacks exploit the fact that ML models often behave differently on the data that they were trained on than they behave on the test data~\cite{shokri2017membership}. 
Specifically, as introduced in~\cite{shokri2017membership}, these attacks utilize shadow models that only require access to the prediction vector of the target ML system. However, these attacks assume the adversary has partial information about the target system's training data by exploiting either partial access to raw \textit{[AC4]} and training data \textit{[AC4]} or by having prior partial knowledge on the training data \textit{[AK5]}.

\noindent \textbf{[AT15] Gradient-based white-box membership inference attacks.}
In these types of attacks, an adversary has partial knowledge on the training data \textit{[AK5]}, \textit{[AC5]}, and \textit{[AC6]}; and complete access \textit{[AC2]},\textit{[AC3]} and knowledge \textit{[AK2]},\textit{[AK3]}, \textit{[AK4]} on the target system's model. Membership inference \textit{[T3]} is performed in these types of attacks based on gradient behavior observed during the training phase~\cite{nasr2019comprehensive}.

\noindent \textbf{[AT16] Shadow training-based property inference attacks.}
In these types of attacks, an adversary extracts information about the features that are not correlated with the learning task \textit{[T4]}.  These attacks can be performed using shadow models, leveraging the adversary's access to the target model's predictions (\textit{[AC9]},\textit{[AC10]}), and partial knowledge of the training data, leveraging \textit{[AC4]}, \textit{[AC5]}, \textit{[AK5]}, and the target system's task \textit{[AK8]}~\cite{ateniese2015hacking,ganju2018property,song2020information}.

\noindent \textbf{[AT17] Maximum a posteriori-based gray-box data reconstruction attacks.}
In these types of attacks~\cite{fredrikson2014privacy,hidano2017model} falling under \textit{[T5]}, an adversary reconstructs training samples, and their respective labels \textit{[A1],[A5]}, of the target system's ML model [A9]. 
The adversary uses a maximum a posteriori (MAP) estimate of the attribute that maximizes the probability of observing the known parameters while assuming partial access \textit{[AC2], [AC3], [AC4], [AC5], [AC9], [AC10]} and knowledge \textit{[AK5], [AK6], [AK8]} of the target model \textit{[A9]} and feature information from the training data.

\noindent \textbf{[AT18] Gradient optimization-based gray-box data reconstruction attacks.}
In these attacks \cite{fredrikson2015model}, an adversary has model knowledge \textit{[AK2]}, task knowledge \textit{[AK8]}, and partial training data knowledge \textit{[AK5]}. 
In addition, the adversary can also obtain the required knowledge by exploiting \textit{[AC2]}, \textit{[AC3]}, \textit{[AC4]}, and \textit{[AC5]}. 
These attacks solve an optimization problem using gradient descent in the input sample space to recover the input data point \textit{[T5]}.

\noindent \textbf{[AT19] Black-box data reconstruction attacks.}
In these types of attacks~\cite{yang2019neural}, an adversary reconstructs training samples \textit{[T5]} while having limited training task information [AK8] and only query access \textit{[AC9]} and \textit{[AC10]} to the target model. These attacks use an autoencoder setting where the target model plays the role of an encoder, and the trainable decoder network tries to reconstruct the training sample on the prediction vector \textit{[AC9]} or \textit{[AC10]} of the target model.

\noindent \textbf{[AT20] Query-based black-box model extraction attacks.}
In these types of attacks~\cite{chandrasekaran2020exploring,juuti2019prada,oh2019towards,papernot2017practical}, an adversary tries to extract complete information of a target ML model \textit{[A9]} by fully reconstructing it or creating a substitute model that behaves very similarly \textit{[T6]}. These attacks require only query access \textit{[AC9],[AC10]} to the target model and limited knowledge of the training task \textit{[AK4],[AK5],[AK8]}.

\noindent \textbf{[AT21] Privacy attacks exploiting the attacker's access to the target system.}
In these types of attacks, an adversary can directly access the target component of the target system to perform a privacy attack \textit{[AG3]}. For example, the training data \textit{[A1]} and \textit{[A5]} in membership inference, data property inference, or data reconstruction attacks is compromised through \textit{[AC4]}, \textit{[AC5]}, and \textit{[AC6]}. In model extraction attacks, the adversary directly accesses and steals the target model \textit{[A9]} through \textit{[AC2]}. 
The adversary gains access to the target system through a conventional cyberattack.

\section{Common Adversarial ML Vulnerability Scoring System (CMLVSS)}
As demonstrated in Section~\ref{sec:security_assessment}, adversarial machine learning (AML) attack techniques may require a different threat model and commonly result in a different security impact.
Thus, it is possible that different attack techniques do not have the same severity level.
Unfortunately, recent works on AML do not provide any method for quantifying the severity of AML attacks. 
The development of such a method is crucial for the ability to integrate AML threats into traditional cybersecurity risk assessment tools.

In this section, we present a method for assigning a severity score to AML attacks, similar to the common vulnerability scoring system (CVSS).
The presented method assigns a severity score to traditional security vulnerabilities. 
We utilize the threat analysis conducted in Section~\ref{sec:security_assessment} to identify the attributes of AML attacks that affect the attack's severity level.
Then, we assist security experts for ranking these attributes, and we use ranking to derive the security severity score for each attack.

One of the first challenges facing when using individuals for ranking multiple criteria is balancing the trade-off between time complexity and accuracy.
Specifically, previous studies have demonstrated that a synchronous pairwise comparison between pairs of criteria has resulted in a very accurate ranking comparing to ranking multiple criteria simultaneously~\cite{saaty2008decision}. 
However, a pairwise comparison is not feasible for a large number of criteria. 
To address this challenge, we utilize a technique from the domain of group decision making coined the analytic hierarchy process (AHP)~\cite{saaty2008decision}, which is specifically designed to moderate these drawbacks. 

\subsection{The Analytic Hierarchy Process (AHP)}
The AHP is a comprehensive framework for quantifying the weights of decision criteria using a group of experts.
According to the AHP, individual expert estimates (using a specially designed questionnaire) the relative magnitudes of attributes concerning the decision criteria through pairwise comparisons.
However, in contrast to naive pairwise comparisons, which are not feasible for a large number of criteria, the AHP suggests decomposing the decision problem into a hierarchy of more easily comprehended sub-problems, each of which can be analyzed independently.
That is, within the AHP, experts only compare pairs of elements at the same level of the hierarchy, thus reducing the amount of comparison needed.
Another advantage of the AHP is the ability to measure the internal consistency of each expert, as well as the agreement among a group of experts (we will elaborate on these measures in the following sections).

\subsection{Using the AHP for Ranking AML Attacks}
The process of ranking AML attacks using the AHP includes the following phases:
\textit{i)} decomposing the decision problem into a hierarchy of sub-problems;
\textit{ii)} constructing the questionnaire;
\textit{iii)} calculating the weights of each attack attributes based on individual experts;
\textit{iv)} validating internal consistency;
\textit{v)} testing external agreement; and
\textit{vi)} calculating the severity score of each attack based on group of experts.
\begin{enumerate}[label=(\roman*)]
\item \textbf{Decomposing the decision problem into a hierarchy of sub-problems.}
To utilize the AHP for ranking AML attacks by their severity, we need to identify the attributes that affect the severity of an attack and organize them in a hierarchical structure.
To do so, we follow the ontology presented by CVSS for the severity of traditional security vulnerabilities and adjust it to the domain of AML attacks based on the threat model described in Section~\ref{sec:threat_model}.

\begin{figure*}[h]
\centering
\includegraphics[width=1\textwidth]{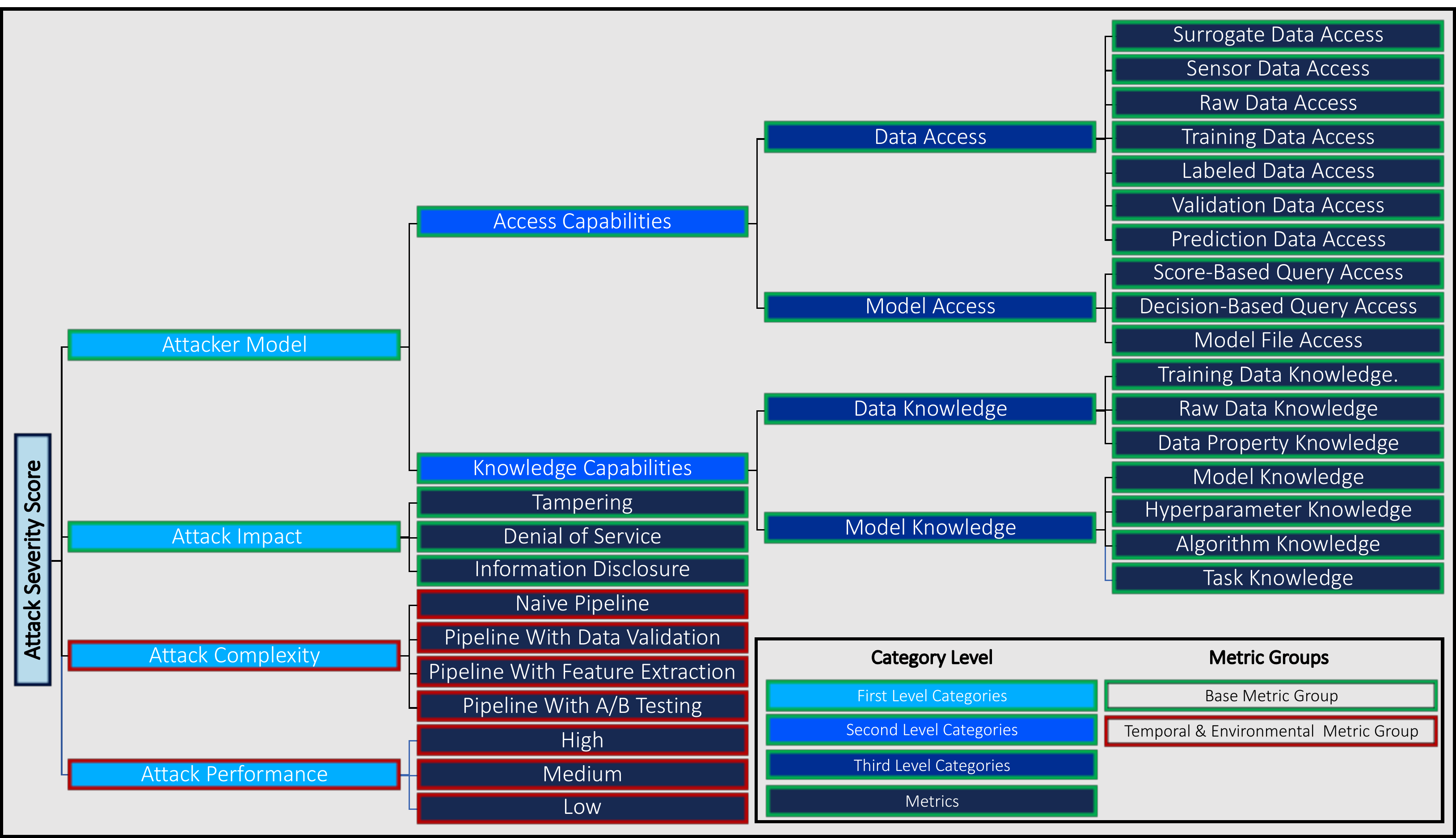}  \caption{Attack severity taxonomy.}
 \label{fig:ahp_taxonomy}
\end{figure*}

The resulted taxonomy, which we refer to as AML attack severity taxonomy, is presented in Figure~\ref{fig:ahp_taxonomy}.
As can be seen, the proposed severity score is composed of four top-level categories: attacker model, attack impact, attack complexity, and attack performance.
Note that similar to the CVSS, we distinguish between two types of attributes (denoted as metric groups): the base metric group, which represents the intrinsic qualities of an AML attack technique that are constant and do not depend on a specific attack implementation nor a specific environmental configuration; and the temporal \& environmental metric group, which represents the characteristics of a AML attack technique that are unique to a specific environmental configuration or attack implementation.
Then, following the threat model described in Section~\ref{sec:threat_model}, we define two second-level categories for the attacker model category: \textit{attacker access capabilities} and \textit{attacker knowledge capabilities}.
Next, we define third-level categories to distinguish between \textit{data access attributes} and \textit{model access attributes}; as well as to distinguish between \textit{data knowledge} and \textit{model knowledge}.
Finally, the leaves of the taxonomy (i.e., the basic attack attributes) are defined based on the threat model described in Section~\ref{sec:threat_model}.

\item \textbf{Constructing the questionnaire.} In this step, we design a questionnaire that can be completed by security experts for ranking pairs of attributes that reside at the same level within the AML attack severity taxonomy.
The proposed questionnaire was implemented in Excel and provided as supplementary material. 
We also provide screenshots of the questionnaire in Figure~\ref{fig:q_screenshots}.
An example for a comparison question is: \textit{Which of the following two capabilities (denoted by "A" and "B") is more difficult to obtain by an attacker, and to what extent?}
where A = \textit{Surrogate Data Access} , and B = \textit{Sensor Data Access}. The possible answers for every question are presented in a nine-level Likert scale (for importance ranking).

\begin{figure}
\begin{subfigure}{1\textwidth}
  \centering
  % include first image
  \includegraphics[width=1\linewidth]{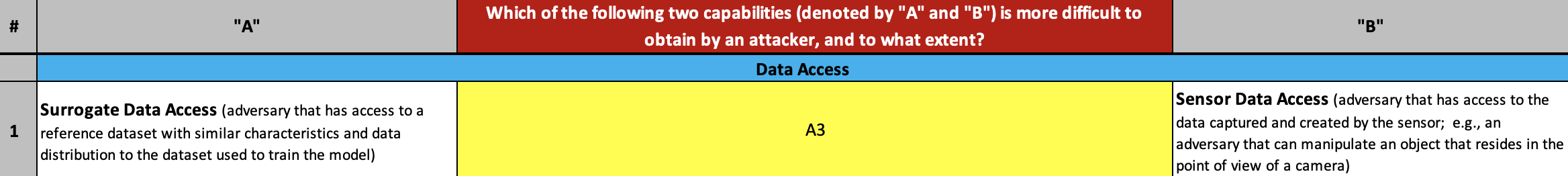}  
  \caption{Ranking attributes (metrics)}
  \label{fig:q_metric_level}
\end{subfigure}
\begin{subfigure}{1\textwidth}
  \centering
  % include second image
  \includegraphics[width=1\linewidth]{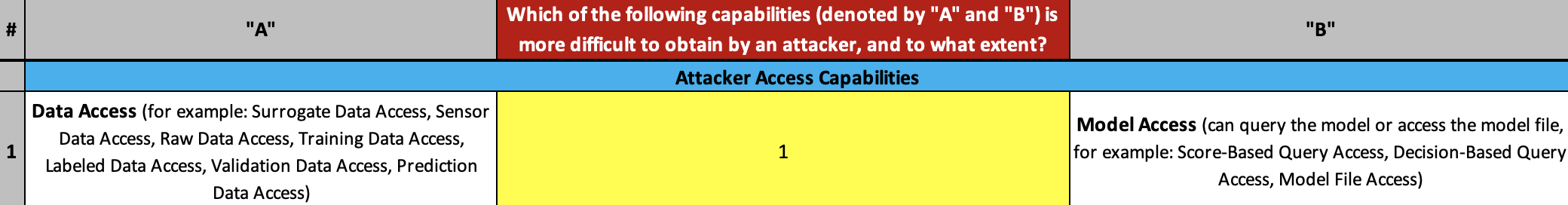}  
  \caption{Ranking third-level categories}
  \label{fig:q_third_level}
\end{subfigure}
\begin{subfigure}{1\textwidth}
  \centering
  % include third image
  \includegraphics[width=1\linewidth]{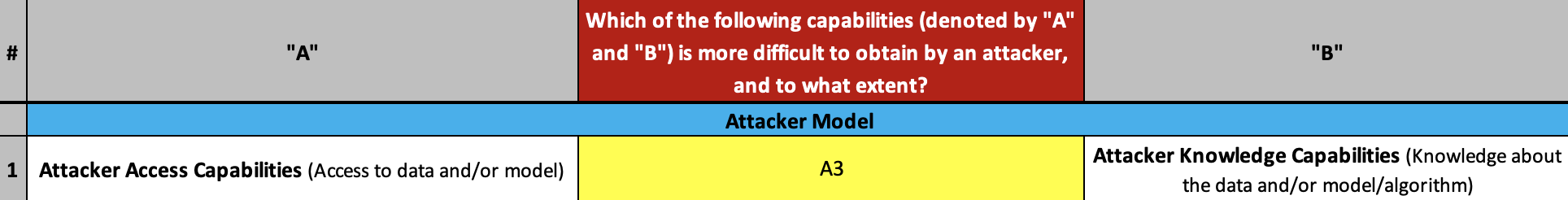}  
  \caption{Ranking second-level categories}
  \label{fig:q_second_level}
\end{subfigure}
\begin{subfigure}{1\textwidth}
  \centering
  % include fourth image
  \includegraphics[width=1\linewidth]{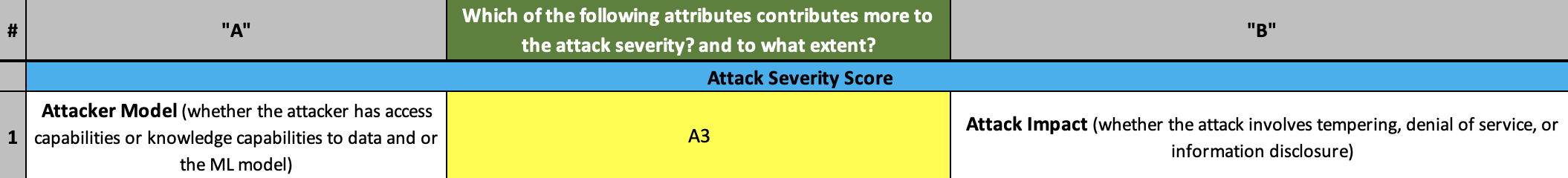}  
  \caption{Ranking first-level categories}
  \label{fig:q_first_level}
\end{subfigure}
\caption{Screenshots from the designed questionnaire.}
\label{fig:q_screenshots}
\end{figure}

\item \textbf{Calculating the weights of each attack attributes based on individual experts.}
In this step, the score for each element in the taxonomy is calculated according to the AHP methodology~\cite{saaty2008decision}.
This is done for each ranker individually.
    
\item \textbf{Validating internal consistency.}
In this step, we validate the internal consistency of each ranker.
The validation is performed dynamically during ranking, using the consistency ratio measure (denoted by CR).
According to the AHP, CR values that are less than 0.1 are considered consistent.
Therefore, if the consistency ratio measure exceeds 0.1, the experts were requested to update their ranking to preserve consistency.
By the end of this phase, all experts achieved a CR values that are less than 0.1 for all comparison levels.

\item \textbf{Testing external agreement.}
In this phase, we tested the external agreement among different experts.
The evaluation was performed using Kendall's W test (i.e., Kendall's coefficient of concordance), which is a non-parametric statistical test commonly used for assessing agreement among raters.
Kendall's W ranges from 0 (no agreement) to 1 (complete agreement), where a value greater than 0.6 is considered a strong agreement.

\item \textbf{Calculating the severity score of each attack.}
Finally, we compute the average score for each element in the taxonomy across all experts.
At the end of this step, each element (including the leaves which represent the threat model criteria) are assigned a score that represents the contribution (weight) of each criterion to the impact of the corresponding AML attack.
In this final step, given an AML attack, we can derive the severity score of the attack by integrating (weighted average) the scores of all criteria that define the attack. 
\end{enumerate}

\subsection{Results}
We let eight experts in the domains of AML and cyber security filling the questionnaire.
When testing the external agreement among the different experts, we achieved an agreement factor of 0.75 indicating a strong agreement.
Therefore, the scores of each element in the taxonomy was computed using the questionnaires of all raters.

\begin{figure*}[t]
            \centering
            \includegraphics[width=1\textwidth]{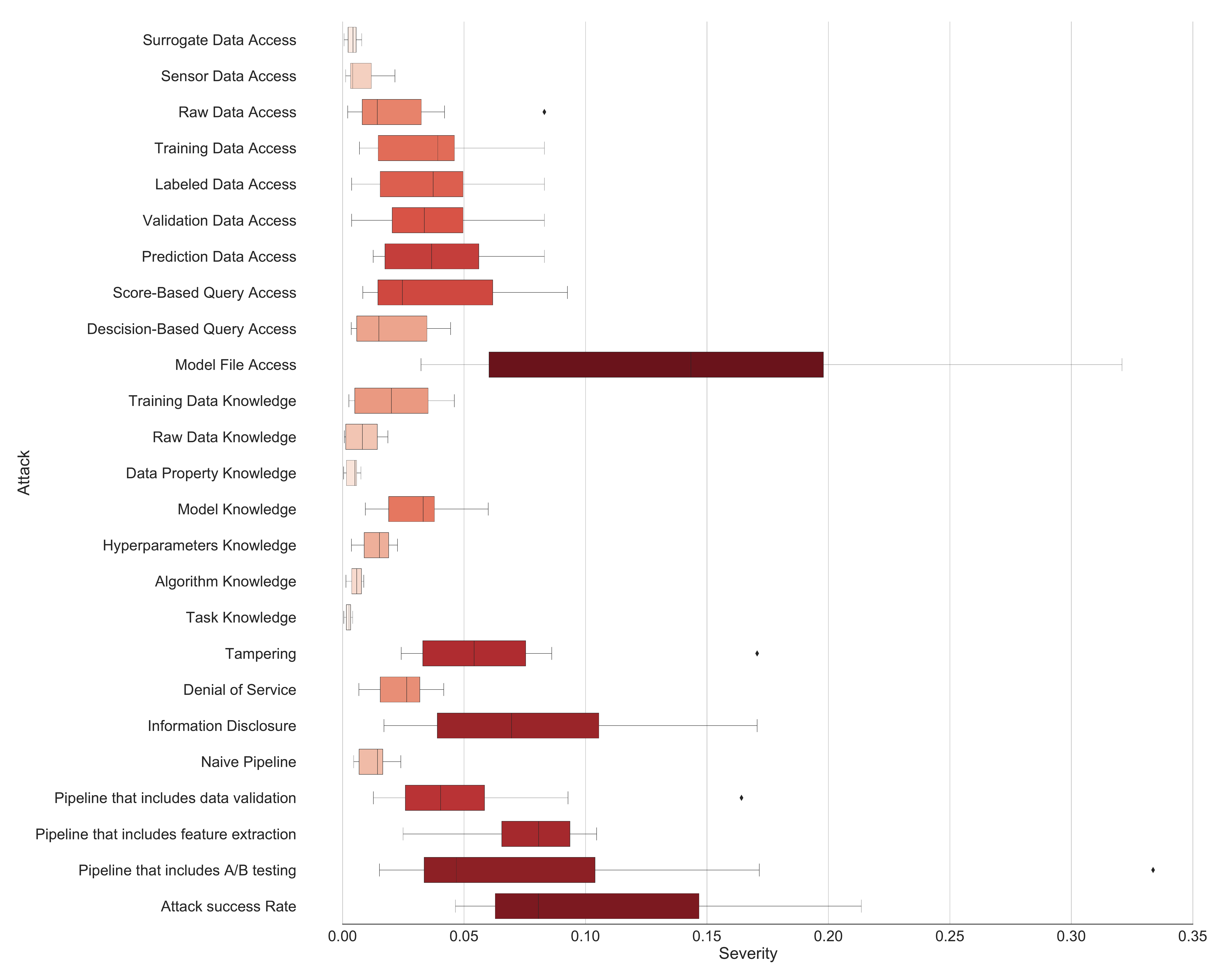}
             \caption{Experts' ranking results: metric level.}
            \label{fig:metric_ranking}
\end{figure*}

In Figure~\ref{fig:metric_ranking} we present the importance of each metric on the attack severity level. 
Note that the importance of attributes that are associated with the attacker model or attack complexity categories quantifies the \textit{negative} impact of the attribute on the severity of the attack. 
On the other hand, the importance of attributes that are associated with the attack impact and attack success rate indicates the \textit{positive} impact on the severity of the attack.

As can be seen, in terms of the attacker model, access to the model file (that is currently deployed) is considered as the capability that is most difficult to obtain by an attacker.
In addition, access to training data, labeled data, validation data, and prediction data, is also considered difficult to obtain by an attacker.
On the other hand, access to surrogate data and sensor data is considered easy to obtain by an attacker.
Unsurprisingly, obtaining data/model knowledge is easier than obtaining data/model access.

Furthermore, the results show that attacks that resulted in tampering or information disclosure are more severe than attacks that resulted in denial of service, and that it is far more difficult to implement an attack on deployments that include A/B testing and feature extraction.

\begin{figure}[h]
            \centering
            \includegraphics[width=1.0\textwidth]{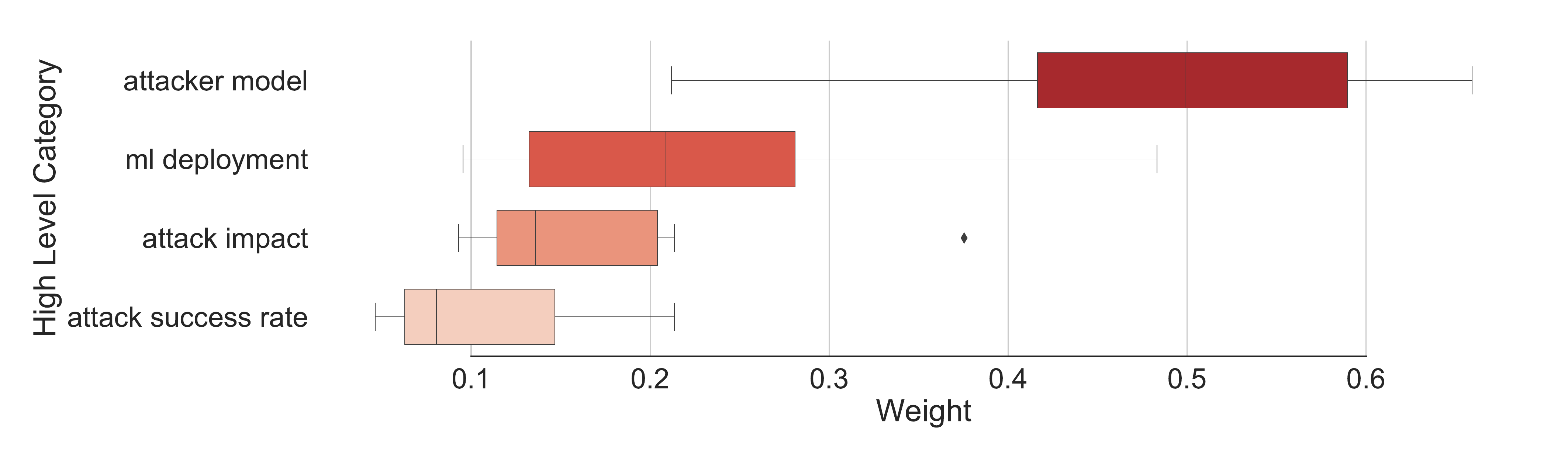}
             \caption{Experts' ranking results: High-level category.}
            \label{fig:high_level_category_ranking}
\end{figure}

In Figure~\ref{fig:high_level_category_ranking} and in Table~\ref{tab:attack_techniques} we analyzed the importance of high-level categories on the attack severity. 
As can be seen, the attacker model is considered as the most important aspect of attack severity, and that attack impact and attack success rate are less important. 

\begin{table*}[h!]
\tiny
\centering
\hspace*{-0.4cm}
\begin{tabular}{@{}
|m{0.018\textwidth}
m{0.018\textwidth}
m{0.018\textwidth}
m{0.018\textwidth}
m{0.018\textwidth}
m{0.018\textwidth}
m{0.018\textwidth}
m{0.018\textwidth}
m{0.018\textwidth}
m{0.018\textwidth}
m{0.018\textwidth}"
m{0.018\textwidth}
m{0.018\textwidth}
m{0.018\textwidth}
m{0.018\textwidth}
m{0.018\textwidth}
m{0.018\textwidth}
m{0.018\textwidth}
m{0.018\textwidth}"
m{0.018\textwidth}
m{0.018\textwidth}
m{0.018\textwidth}"
m{0.018\textwidth}
m{0.018\textwidth}
m{0.018\textwidth}
m{0.018\textwidth}|
@{}}
\Xhline{3\arrayrulewidth}

\multicolumn{19}{|c"}{\underline{Threat Model}} & 
\multicolumn{3}{c"}{\makecell{Attack \\ Impact}} &

\multicolumn{4}{c|}{\makecell{Environmental \\ Metrics}} \\
\multicolumn{11}{|c}{Access Capabilities} &
\multicolumn{8}{c"}{System Knowledge} & & & & & & &\\ \Xhline{3\arrayrulewidth}

\rotatebox{90}{Pipeline Access $[AC1]$} & 
\rotatebox{90}{Model Access $[AC2]$} & 
\rotatebox{90}{Prediction Access $[AC3]$} & 
\rotatebox{90}{Raw Data Access $[AC4]$} &
\rotatebox{90}{Training Data Access $[AC5]$} &
\rotatebox{90}{Labeled Data Access $[AC6]$} &
\rotatebox{90}{Validation Data Access $[AC7]$} &
\rotatebox{90}{Surrogate Data Access $[AC8]$} &
\rotatebox{90}{Score-Based Query Access $[AC9]$} &
\rotatebox{90}{Decision-Based Query Access $[AC10]$} &
\rotatebox{90}{Sensor Data Access $[AC11]$} &

\rotatebox{90}{Perfect Knowledge $[AK1]$} &
\rotatebox{90}{Model Knowledge $[AK2]$} &
\rotatebox{90}{Hyperparameter Knowledge $[AK3]$} &
\rotatebox{90}{Algorithm Knowledge $[AK4]$} &
\rotatebox{90}{Training Data Knowledge $[AK5]$} &
\rotatebox{90}{Raw Data Knowledge $[AK6]$} &
\rotatebox{90}{Data Property Knowledge $[AK7]$} &
\rotatebox{90}{Task Knowledge $[AK8]$} &

\rotatebox{90}{Tampering $[AG1]$} & \rotatebox{90}{Denial of Service $[AG2]$} &
\rotatebox{90}{Information Disclosure $[AG3]$} &

\rotatebox{90}{Naive Pipeline} &
\rotatebox{90}{Pipeline With Data Validation} &
\rotatebox{90}{Pipeline With Feature Extraction} &
\rotatebox{90}{Pipeline With A/B Testing}

\\ \Xhline{3\arrayrulewidth}

\cellcolor{Grey} $\Sigma_{ac}$ &
\cellcolor{VeryLow} .144 &
\cellcolor{Medium} .040 &
\cellcolor{Medium} .025 &
\cellcolor{Medium} .037 &
\cellcolor{Medium} .038 &
\cellcolor{Medium} .038 &
\cellcolor{VeryHigh} .004 &
\cellcolor{Medium} .039 &
\cellcolor{Medium} .020 &
\cellcolor{VeryHigh} .008 &
\cellcolor{Grey} $\Sigma_{sk}$ &
\cellcolor{Medium} .031 &
\cellcolor{Medium} .014 &
\cellcolor{VeryHigh} .006 &
\cellcolor{Medium} .022 &
\cellcolor{Medium} .025 &
\cellcolor{VeryHigh} .004 &
\cellcolor{VeryHigh} .002 &
\cellcolor{Grey} .066 &
\cellcolor{Grey} .024 &
\cellcolor{Grey} .081 &
\cellcolor{Medium} .013 &
\cellcolor{VeryLow} .055 &
\cellcolor{VeryLow} .075 &
\cellcolor{VeryLow} .095 \\  \Xhline{3\arrayrulewidth}

\end{tabular}%
\caption{Adversarial machine learning attack techniques.}
\label{tab:attack_techniques}
\end{table*}

\begin{figure*}[h]
            \centering
            \includegraphics[width=1\textwidth]{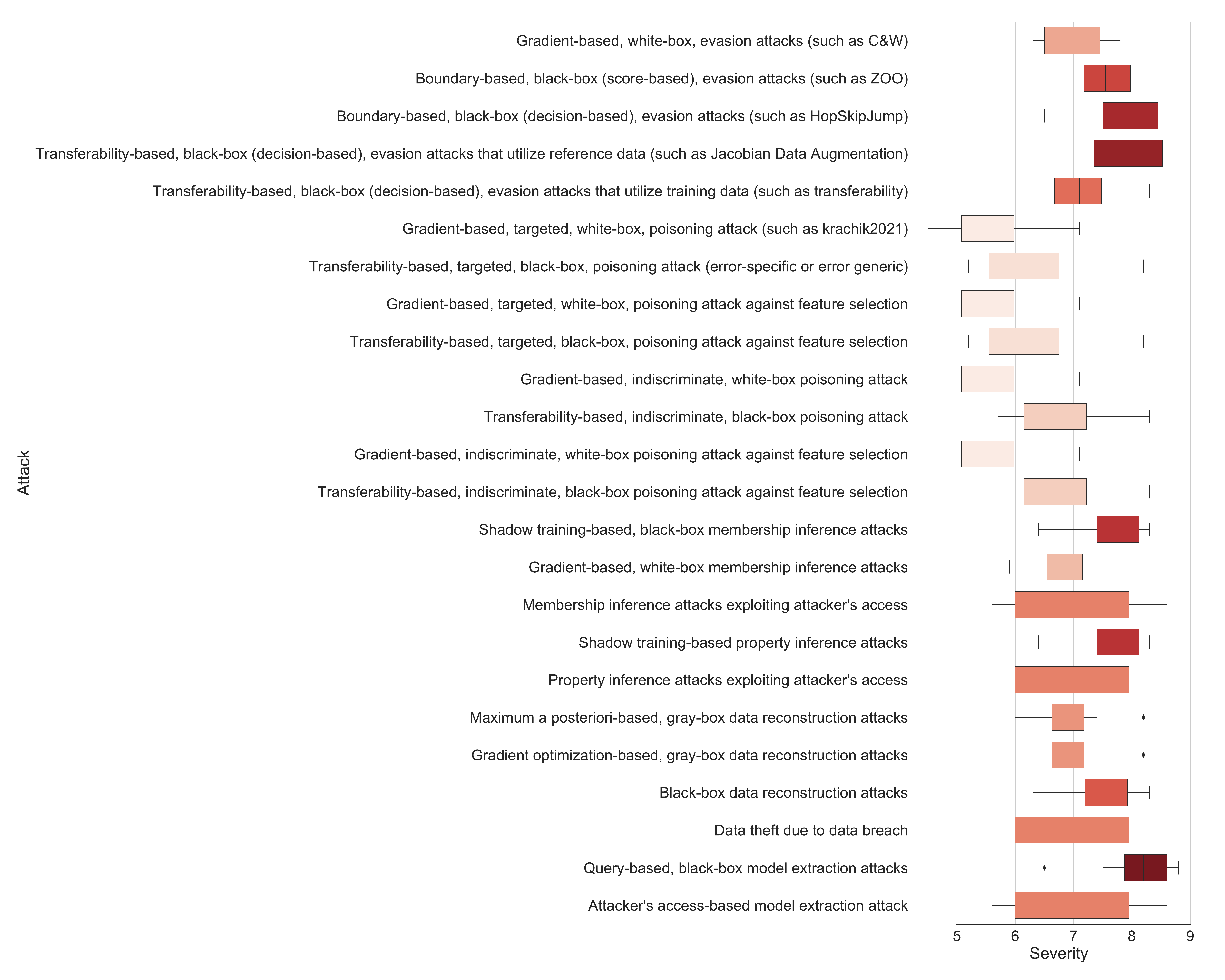}
             \caption{Experts' ranking results: Attack ranking.}
            \label{fig:attack_ranking}
\end{figure*}

In Figure~\ref{fig:attack_ranking} we present the severity of the different attack techniques. 
As can be seen, the most severe attack techniques are boundary-based, black-box (decision-based), evasion attacks (such as HopSkipJump), transferability-based, black-box (decision-based), evasion attacks that utilize reference data (such as Jacobian Data Augmentation), and query-based, black-box model extraction attacks. 
We attribute that to the fact that these attacks can operate in a complete black-box setting with minimal demands from the attacker.
On the other hand, poisoning attacks received relatively low severity score (despite their impact). 
We attribute that to the fact that poisoning attacks require the attacker to have the ability to write into the training set.

\section{MulVAL Extension \label{sec:mulval_extension}}

\subsection{Introduction to Logical Attack Graphs}
A logical attack graph is a directed graph that represents all possible attack scenarios.
It specifies the relations between the specific system configuration (e.g., running services, installed software, and existing security vulnerabilities) and the attacker’s potential privileges (e.g., executing arbitrary code, denial of service, and privilege escalation).
To generate a logical attack graph, the consequence of each attack phase should be expressible as a propositional formula of the system configurations. 
In logical attack graph terminology, propositional formulas are referred to as \textit{interaction rules}, and system configurations (and attacker's privileges) are referred to as \textit{predicates} (which can be either \textit{primitive}, i.e., a basic standalone fact, or \textit{derived}, i.e., complex facts that are derived based on other facts using an interaction rule).
%A formal definition of a logical attack graph is presented in Appendix~\ref{app:logical_attack_graph}. 

Formally, a logical attack graph is defined as a tuple $AG = (N_r, N_p, N_d, E, \mathcal{L}, \mathcal{G})$, where:
\begin{itemize}%[label=\roman*]
    \item $N_r$ - The set of derivation nodes. These nodes correspond to interaction rules and imply an \textit{AND} relation between their incoming nodes.
    \item $N_p$ - The set of nodes that represent primitive facts.
    \item $N_d$ - The set of nodes that represent derived facts. These nodes imply an \textit{OR} relation between their incoming nodes.
    \item $E \subseteq \{(N_p \cup N_d)\times N_r\} \cup \{N_r \times N_d\}$ - The set of edges.
    \item $\mathcal{L}$ - A mapping between nodes and their labels.
    \item $\mathcal{G}$ - The node that represents the attacker's goal.
\end{itemize}
%\subsection{Interaction Rule Written in Datalog \label{app:simple_interaction_rule}}

The main benefits of a risk assessment based on logical attack graphs are twofold: First, while traditional risk assessment methods analyze each vulnerability individually, attack graph-based risk assessment also models the interactions between vulnerabilities and the lateral movements of the attacker. 
Second, while traditional risk assessment methods evaluate the severity of each vulnerability regardless of the specific environmental deployment and settings, attack graph-based risk assessment considers the specific preconditions and impact of exploiting security vulnerabilities on the specific target environment. 
Thus, attack graphs enable a very accurate and concrete risk assessment.

\subsection{Overview of The MulVAL Framework}
The MulVAL framework is an open-source tool for generating and analyzing logical attack graphs~\cite{ou2005mulval}.
MulVAL has three main benefits in comparison to previously suggested tools:
\begin{enumerate}[label=(\roman*)]
\item \textbf{Automation.} MulVAL models the interactions of software vulnerabilities with system and network configurations, while \textit{automatically} extracting information from formal vulnerability databases (e.g., the National Vulnerability Database~\cite{nvd}) and network scanning tools (e.g., Nessus~\cite{nessus}).
\item \textbf{Efficiency.} The complexity of other existing tools (e.g., model checking tools) when enumerating all possible attack paths is exponential in the size of the input~\cite{ou2005mulval}. 
In contrast, MulVAl enumerates all possible attack paths in a polynomial time. 
\item \textbf{Extendibility.} MulVAL is an open-source framework, therefore it can be extended with new features. Furthermore, MulVAL's expressive capability can be extended with new attack scenarios.
\end{enumerate}

Consequently, multiple risk assessment and countermeasure planning algorithms use MulVAL as a method for estimating risk~\cite{stan2021heuristic,binyamini2020automated,agmon2019deployment,gonda2018analysis}.

Within MulVAL, interaction rules and predicates are written in Datalog, which is a subset of the Prolog programming language. 
The Datalog language includes three main entities, namely the variables, constant values, and predicates.
Variables always start with an uppercase letter; they can be instantiated with any value during evaluation.
Constant values always start with a lowercase letter, and they are set during the formulation of the rule. 
Predicates are an atomic formula of the form: $p(t_1,...,t_k)$, where each argument $t_i$ can be either a variable or a constant value.
%An example of a simple interaction rule written in Datalog is presented in Listing~\ref{li:simple_interaction_rule}.

In Listing \ref{listing:example}, we present a simple interaction rule (written in Datalog) that represents a general attack technique -- the remote exploit of a privilege escalation vulnerability in a program -- which leads to the attacker’s potential privilege to execute arbitrary code on the host running the program.

\lstinputlisting[
  style      = Prolog-pygsty,
  caption    = {An example to a simple interaction rule.},
  label = {listing:example},
  float=h,
  floatplacement=tbp
]{example.pl}

This generic interaction rule specifies the preconditions and consequence for this attack.
Specifically, if a $Program$ is running in a $Privilege$ on a $Host$ as a network service listening on $Protocol$ and $Port$ (the first precondition); and this $Program$ contains a remotely exploitable ($remote$) vulnerability ($VulID$) whose impact is privilege escalation (the second precondition); and the $Attacker$ can access the service running on $Host$ through the network on $Protocol$ and $Port$ (the third precondition), then the attacker can execute arbitrary code on the $Host$ that $Privilege$ (attack consequence).

\subsection{Limitations of MulVAL}
While MulVAL is a very powerful tool, like all risk assessment frameworks it requires high maintenance.
Specifically, to cope with new attack trends and emerging threats, risk assessment frameworks must be updated regularly with up-to-date information on new vulnerabilities and attack techniques.
In MulVAL, attack techniques are represented using predicates and interaction rules, and the process of formulating new predicates and interaction rules for representing a new attack technique is referred to as \emph{attack modeling}.
Consequently, to incorporate attacks on ML systems, a wide range of new attack techniques (such as adversarial ML) should be considered and modeled.

Previous attempts to extend MulVAL have focused on network attacks~\cite{bacic2006mulval,froh2009mulval,stan2019extending,mavani2017modeling,liu2015logic,acosta2016augmenting}, data vulnerability, cloud security~\cite{el2017game}, and IoT environments~\cite{agmon2019deployment}.
These extensions, however, do not consider attacks on ML systems.
For instance, in a \textit{poisoning attack}, the attacker must have the ability to manipulate the model's training data, however the relation between a model and its training set is a property that is specific to ML applications.
This property is not considered in the previous extensions to MulVAL. 
Therefore, existing methods cannot be used to quantify the risk of an enterprise network that includes ML systems.
To the best of our knowledge, a threat analysis framework that considers cyber attacks on ML systems has yet to be presented. 

In this work, we follow the methodology presented in~\cite{inokuchi2019design} and extend MulVAL to represent cyberattacks on ML systems.
The proposed extension enhances the power of attack graph-based risk assessment by introducing paradigms that have not been modeled before, thus providing security practitioners with a tactical tool for evaluating the impact and quantifying the risk of a cyberattack targeting ML systems.

%The MulVAL framework is an open-source tool for generating and analyzing logical attack graphs~\cite{ou2005mulval}.
%While MulVAL is a very powerful tool, like all risk assessment frameworks it requires high maintenance.  
%Specifically, to cope with new attack trends and emerging threats, risk  assessment frameworks must be updated regularly with up-to-date information on new vulnerabilities and attack techniques.
%Consequently, to incorporate attacks on ML systems, a wide range of new attack techniques (such as AML) should be considered and modeled.

\subsection{The Proposed Extension}
In this section, we present a MulVAL extension that considers attacks on ML systems. 
Due to space limitations, we are unable to present all of the predicates and interaction rules included in the proposed extension. 
Instead, we provide guidelines for creating these predicates and demonstrate the guidelines using examples. 
The extension (including all of the predicates and interaction rules) is available \href{https://drive.google.com/drive/folders/1vYKeSObPcjJoUgOw9BOngepzZ9ytay-R?usp=sharing}{\textcolor{blue}{\textbf{here}}}.

\noindent\textbf{Modeling pipeline components.}
To incorporate attacks on ML systems within the MulVAL attack graph analysis tool, we must first model the basic components of an ML system.
This is done by creating new \textit{primitive} predicates for all of the target assets described in Section~\ref{subsec:target_assets}.
For example, the most basic components of any ML system are the learning algorithm and the ML model, and in Listing \ref{listing:components_and_data} we demonstrate how we model these components using the Datalog language.
Specifically, $algorithm3$ (line 1) is a \textit{primitive} predicate that specifies a learning algorithm. 
This predicate includes the following two characteristics: $AlgorithmID$ and $AlgorithmHost$, which indicate the path to the source code of the algorithm ($AlgortihmID$) and the IP address of the host storing that source code ($AlgorithmHost$) respectively; $model6$ (line 2) is a \textit{primitive} predicate that specifies an ML model ($ModelID$) that is assigned to a specific pipeline ($PipelineID$) and is created by applying the specific learning algorithm ($AlgorithmID$) on specific training data ($TrainingDataID$).

Following the definition of a learning algorithm and ML model, we define the attributes of such models.
For example, $vulModel5$ (line 3) specifies whether a certain ML model ($ModelID$), trained by a specific learning algorithm ($AlgorithmID$), has a particular type ($Type$) of security vulnerability; and $publicModel$ (line 4) specifies whether a certain model ($ModelID$) is publicly available.

\noindent\textbf{Modeling data types.} 
ML production pipelines have multiple data types. 
In our model, we consider the following nine data types: raw, feature, label, training, validation, evaluation, prediction, and label data.
For each data type, we created a \textit{primitive} predicate that includes all relevant characteristics of that type.

For example, in Listing~\ref{listing:components_and_data} we demonstrate how we model training data (line 5).
Specifically, the $trainingData3$ \textit{primitive} predicate includes the following four characteristics: $pipelineID$, $ModelID$, $DataID$, and $Host$, which specify the pipeline and model that utilizes this data, a unique identifier of that data (e.g., the path of this data), and the location of that data (e.g., the IP of the host storing the data).

\lstinputlisting[
  style      = Prolog-pygsty,
  caption    = {ML components and data types.},
  label = {listing:components_and_data},
  float=h,
  floatplacement=tbp
]{model.pl}

\noindent\textbf{Modeling data operations.}
ML pipelines include multiple data operations, such as data normalization, data aggregation, and feature extraction. 
In our modeling, we refer to these operations as data transformation jobs (line 1 in Listing~\ref{listing:dataTransformation}). 
A data transformation job ($DataTransformationJob$) is a \textit{primitive} predicate that represents a task (identified by $JobID$), which takes data in one format ($InputDataID$) and transforms it to another format ($TransformedDataID$) by applying some transformation (e.g., feature normalization). 
Since the hosts running the task are not necessarily the same as the hosts storing the input/output data, in our modeling, the data transformation job also includes variables that specify the host running the job ($JobHost$), the host storing the input data ($InputDataHost$), and the host storing the transformed data ($TransformedDataHost$).

In addition, since in production pipelines data operations are commonly performed on computing clusters, we extend our modeling to support the use of computing clusters to run data transformation jobs (lines 2-8). 
We start by modeling the relationship between the three basic cluster components: master, edge (gateway host), and worker nodes. 
Specifically, the master node is responsible for task management and resource scheduling among the worker nodes which store data and execute tasks on the data. 
The edge nodes are used for getting data in and out of the cluster and for data processing job submissions. In our modeling, this relationship is represented using the $clusterWorker3$ predicate (line 2), which indicates that a certain host ($WorkerHost$) is a part of a computing cluster for which $ClusterGatewayHost$ and $ClusterMasterHost$ are the gateway and master hosts respectively. 

We further extend the definitions of raw data and data transformation jobs to support the dynamics of a computing cluster. Specifically, we create derivation rules for the $rowData2$ and $dataTransformationJob6$ predicates, which propagate raw data and data transformation jobs from the master node to the worker nodes (line 3-8).

\lstinputlisting[
  style      = Prolog-pygsty,
  caption    = {Data propagation in big data architecture.},
  label = {listing:dataTransformation},
  float=h,
  floatplacement=tbp
]{data-transformation.pl}

\noindent\textbf{Modeling monitoring services.}
To maintain the quality of the deployed model and trigger model retraining or data collection, if necessary, model performance should be monitored continuously. To model monitoring operations we create the $performenceMonitoring$ primitive predicate (line 1, Listing~\ref{listing:monitor}).
We assume that the performance monitoring is performed using some dataset ($DataID$) on a specific worker ($Host$) and concerning a specific pipeline ($PipelineID$). We further assume that the performance monitoring job is controlled by some program ($Program$).
When a decrease in the model's performance is detected, model retraining and/or the collection of new data can be executed. 
We refer to these as $triggerModelTraining4$ and $triggerDataExtraction4$, respectively (lines 2-3). 
To describe the relationship between each ML job and the monitoring job, the worker node monitoring the model performance and starting retraining or data collection is specified ($MonitoringHost$).

\lstinputlisting[
  style      = Prolog-pygsty,
  caption    = {Performance monitoring.},
  label = {listing:monitor},
  float=h,
  floatplacement=tbp
]{monitor.pl}

\noindent\textbf{Modeling attacker's access capabilities.}
We create primitive predicates for all of the access capabilities described in Section~\ref{sec:threat_model}, where each predicate includes the following four characteristics: $Principal$, $PipelineID$, $ModelID$, and $AccessLevel$, which outline the principal that acquires the capability, the affected pipeline, the affected model, and the access level (e.g., read/write, etc.).
Examples of these predicates for model access and query access are presented in Listing \ref{listing:attacker_access_capabilities_predicates} (lines 1-2).

\lstinputlisting[
  style      = Prolog-pygsty,
  caption    = {Attacker's access capabilities.},
  label = {listing:attacker_access_capabilities_predicates},
  float=h,
  floatplacement=tbp
]{attacker_access_capabilities_predicates.pl}

We further create derivation rules that can be used to assess the different access capabilities from the configuration (state) of the target environment.
For example, we create a derivation rule for the $modelAccess4$ predicate, for cases in which a malicious entity can access the model file (lines 3-6). This derivation rule includes the following three preconditions: $malicious1$, which indicates that $Principal$ is a malicious entity; $model6$, which specifies that the ML model is stored in path $ModelID$ in host $ModelHost$; and $accessFile4$, which indicates that the malicious entity ($Principal$) can access the model file stored in path $ModelID$ in host $Host$.

\noindent\textbf{Modeling attacker's knowledge.}
Based on the threat model described in Section~\ref{subsubsec:attacker_knowledge}, we create primitive predicates for all of types of information available to the attacker. 
Each predicate includes the following three characteristics: $Principal$, $PipelineID$, and $ModelID$, which outline the principal that acquires that knowledge on the target system, pipeline, and model.
An example of the predicate created for model knowledge is presented in Listing~\ref{listing:attackerKnowledge_predicates} (line 1). 

We further create derivation rules that can be used to assess system knowledge from the state of the target environment.
For instance, we create derivation rules for the $modelKnowledge3$ predicate, for cases in which an attacker acquires model knowledge through access to the model file (lines 2-5 in Listing~\ref{listing:attackerKnowledge_predicates}).
This derivation rule includes the following three preconditions: $malicious1$, which indicates that $Principal$ is a malicious entity; $model6$, which specifies that the ML model is stored at path $ModelID$ in host $ModelHost$; and $modelAccess4$, which indicates that the malicious entity ($Principal$) can access the model file.
Another example is an attacker that acquires model knowledge through access to public data sources (lines 6-9 in Listing~\ref{listing:attackerKnowledge_predicates}).

\lstinputlisting[
  style      = Prolog-pygsty,
  caption    = {Attacker's knowledge.},
  label = {listing:attackerKnowledge_predicates},
  float=h,
  floatplacement=tbp
]{attacker-knowledge_predicates.pl}

\subsection{Threat Categories and Attack Techniques}
In the subsections above we modeled the basic components of ML systems, the dynamics of such systems, and the various threat models.
In this section we use this modeling to model the different attack techniques and map them to the relevant threat categories.
Specifically, we create derivation rules for each threat category, where each derivation rule represents an attack technique that can be used to materialize the threat and includes the preconditions for executing the attack. 

In Listing~\ref{listing:attacker_actions}, we provide two examples for materializing the evasion threat ($T1$): a boundary-based, black-box, decision-based, evasion attack ($[AT3]$, lines 1-7) and a transferability-based, black-box, poisoning attack ($[AT7]$, lines 7-17).
Specifically, the derivation rule for a boundary-based, black-box, decision-based, evasion attack includes the following five preconditions: $malicious1$, which indicates that $Principal$ is a malicious entity; $model6$, which indicates the existence of an ML model; $vulModel5$, which indicates the existence of an \textit{evasion vulnerability} in the ML model; $queryAccess5$, which specifies that this attack requires query access to the model; and $taskKnowledge$, which indicates that the attacker must know the learning task to successfully execute the attack.

\lstinputlisting[
  style      = Prolog-pygsty,
  caption    = {Threat categories and attack techniques.},
  label = {listing:attacker_actions},
  float=h,
  floatplacement=tbp
]{attacks.pl}

On the other hand, the derivation rule for a transferability-based, black-box, poisoning attack (which can also materialize the evasion threat) includes 10 preconditions. The first five preconditions (lines 8-12) indicate that the adversary can generate a surrogate model and use that model to create poisoning examples that are transferable to the target model; specifically: $malicious1$, which indicates that $Principal$ is a malicious entity; $model6$, which indicates the existence of an ML model; $vulModel5$ which indicates the existence of a \textit{transferability vulnerability} in the ML model;
$taskKnowledge$, which indicates that the attacker must know the learning task to successfully execute the attack; and $surogateDataAccess4$, which specifies that the adversary has access to a surrogate dataset with similar attributes and statistical properties.
The next four preconditions (lines 13-17) indicate that the adversary can inject the generated poisoning examples into the training set of the target model, specifically: $vulModel5$, which indicates the existence of a \textit{poisoning vulnerability} in the target ML model; $trainingDataAccess6$ and $labeledDataAccess6$, which specify that the adversary (denoted by $Principal$) can write to the model's training dataset; and $modelRetrainedContinously2$, which indicates that the target model will eventually be trained on the poisoned training data.
The last precondition (line 17) indicates that the adversary has limited query access to the target model. This precondition is important so the attacker can send the crafted samples to the model for classification. 

\subsection{Risk Assessment Using Attack Graphs and the CMLVSS \label{sec:attack_graph_risk}}
In this section we estimate the risk of each attack path using the proposed AML attack scoring system. 
Our method is based on the risk calculation presented in~\cite{stan2021heuristic} with adjustment to the domain of AML attacks.
Specifically, the risk equation of a node $a$ is defined as follows:
\begin{equation}
    Risk(a) = I_a \cdot LH(a)
\end{equation}
where $I_a$ is the \textit{impact} of exploiting the compromised asset represented by the node $a$, and $LH$ is the likelihood of the attacker to reach that asset.
Note, for AML threats, the impact is defined based on the attack impact metric from CMLVSS.

The likelihood equations (i.e., $LH({a})$) are computed for each node $a$ in the attack graph and are defined recursively by Equations \eqref{eq:vul-risk_aml}, \eqref{eq:and-prob} and \eqref{eq:or-prob}. 
Concretely, if the node $a$ represents a traditional vulnerability, $LH(a)$ is calculated according to the exploitability metrics of the CVSS v.2 base score, which is commonly used to assess security risk.
Specifically, we assign the probability of vulnerability exploitation based on the CVSS v.2 access complexity metric, as specified in Equation \eqref{eq:vul-risk_tv}. 

\begin{equation}
    % \fontsize{7.5pt}{8.5pt}\selectfont
    LH(a \in TV) =
    \begin{cases}
        0.35, & AC=High \\
        0.61, & AC=Medium \\
        0.71, & AC=Low
    \end{cases}
    \label{eq:vul-risk_tv}
\end{equation}

\noindent where $TV$ denotes traditional vulnerability and $AC$ denotes the CVSS v.2 access complexity metric.
We focus only on this metric, because it represents information that cannot be expressed by the other attack graph nodes, unlike the other exploitability metrics (i.e., access vector and authentication).

On the other hand, if the node $a$ represents AML vulnerability,
$LH(a)$ is calculated according to the attack performance metric from CMLVSS.
Specifically, we assign the probability of AML vulnerability exploitation as specified in Equation \eqref{eq:vul-risk_aml}. 

\begin{equation}
    % \fontsize{7.5pt}{8.5pt}\selectfont
    LH(a \in TV) =
    \begin{cases}
        0.35, & AP=High \\
        0.61, & AP=Medium \\
        0.71, & AP=Low
    \end{cases}
    \label{eq:vul-risk_aml}
\end{equation}

Note, if the node $a$ does not represent a vulnerability, then $LH(a)=1$ (i.e., any other fact can be materialized in any situation).

The following equations describe the likelihood calculation of \textit{AND} and \textit{OR} nodes respectively. Note that for an \textit{OR} node to be materialized, it is sufficient if one of its parents is materialized. 
Therefore, in order to consider each parent only once in the likelihood computation, we use the inclusion-exclusion principle.

\begin{equation}
    % \fontsize{7.5pt}{8.5pt}\selectfont
    LH(AND) = \prod_{a \in AND_{in}} LH(a)
    \label{eq:and-prob}
\end{equation}

\begin{equation}
    % \fontsize{7.5pt}{8.5pt}\selectfont
    \begin{split}
    & LH(OR) = \left(\sum_{i=1}^{R} LH(a_i) - \right. \\
    & \left. \sum_{i_1<i_2} LH(a_{i_1}) \cdot  LH(a_{i_2}) + \cdots + \right. \\
    & \left. (-1)^{R} \cdot \sum_{i_1<\cdots<i_{R-1}} LH(a_{i_1})\cdot ... \cdot LH(a_{i_{R-1}}) +  \right. \\ 
    & \left. (-1)^{R+1}\cdot \prod_{i=1}^{R} LH(a_i) \right) \\
    & \\
    & where ~R = |OR_{in}|~and~OR_{in} = \{a_1,a_2,...,a_R\}.
    \end{split}
    \label{eq:or-prob}
\end{equation}

Algorithm \ref{alg:build-risk-eq} describes the process of building the likelihood equations for each node in a given attack graph (denoted by $AG$). It maintains two sets of nodes: a $processed$ set that contains all of the nodes that have already been processed (i.e., their equation has been constructed) and an $unprocessed$ set which contains the remaining nodes. In each iteration, a node that can be processed (i.e., all of its parents have been processed) is chosen, and its representative equation is constructed, until all nodes have been processed.

\begin{algorithm}[t]
	\caption{Build likelihood equations.}
 	%\small
	\label{alg:build-risk-eq}
	\begin{algorithmic}[1]
        \State $processed \gets \{(a,LH(a)) | a \in AG_{LEAF}\}$
        \State $unprocessed \gets AG \setminus AG_{LEAF}$
        \While{$!unprocessed.isEmpty()$}
            \While{$!unprocessed.hasProcessableNode()$}
                \State $a \gets unprocessed.pop()$
                \State $a.updateMA()$
                \If{$\forall k \in a_{in}, k \in processed$}
                    \State $processed \gets processed \cup \{(a,LH(a))\}$
                \Else
                    \State $unprocessed.push(a)$
                \EndIf
            \EndWhile
%            \State $removeCycles(AG)$
        \EndWhile
	\end{algorithmic}
\end{algorithm}

\section{Demonstration \label{sec:demostration}}
In this section, we demonstrate the impact of attack graph analysis using the proposed extension.
We begin by describing a simplified (and vulnerable) target environment that includes ML components. 
Then, we use the proposed extension to generate an attack graph for the target environment and discuss the results.

\subsection{Target Environment}
\begin{figure}[h]
            \centering
            \includegraphics[width=1.0\textwidth]{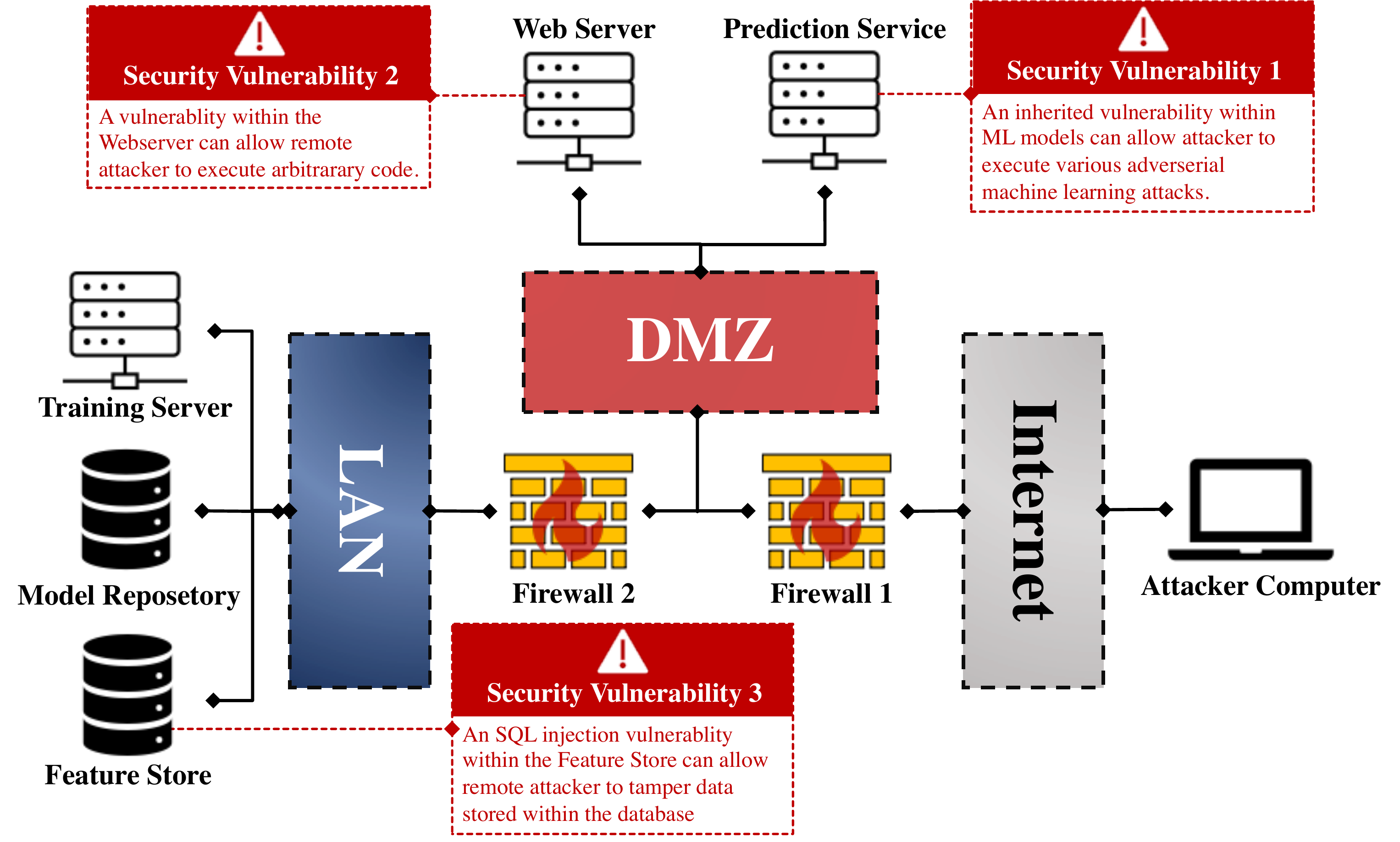}
             \caption{The evaluation framework.}
            \label{fig:evaluation_framework}
\end{figure}

The target environment includes the following eight components (see Figure~\ref{fig:evaluation_framework}): web server, prediction service, training server, model repository, feature store, client application, firewall 1, and firewall 2. 

\begin{enumerate}[label=(\roman*)]
\item \textbf{Web server.} This component is responsible for hosting the enterprise website. 
The web server is located in the DMZ and is reachable from the Internet. 
The web server has access to the feature store, which is located in the LAN. 
The web server runs on a vulnerable version of the Apache web server, which can allow a remote attacker to execute arbitrary code.
    
\item \textbf{Prediction service.} This component is responsible for serving the ML applications that are running on the client application.
The prediction service is located in the DMZ and is reachable from the Internet on TCP port 80. 
The  prediction service has access to the feature store (which is located in the LAN) on port 3036. 
The ML model, which is encapsulated by the prediction service, is vulnerable to evasion and poisoning attacks.
    
\item \textbf{Training server.} This component is responsible for producing the ML model by running the learning algorithms on the training data (from the feature store). 
To cope with concept drifts, an ML model trained on fresh data (which exists in the feature store) is deployed to the serving server every hour (for simplicity, we do not include A/B testing in this environment). 
The training server is located in the LAN. 
    
\item \textbf{Model repository.} This component is responsible for storing trained models, along with their performance and versioning, and other configuration information. The model repository is located in the LAN.

\item \textbf{Feature store.} This component is responsible for storing and logging features for ML tasks. The feature store is located in the LAN and listens for incoming requests On TCP port 3036. 
The feature store is vulnerable to SQL injection, which can allow a remote attacker to tamper with data stored in the feature store (e.g., manipulating the data used to train the ML model).

\item \textbf{Client application.} This component is responsible for querying the prediction service. 
The client application is located on the Internet, and we assume that it can be compromised by an attacker that is masquerading as a legitimate client.

\item \textbf{Firewall 1.} This component is responsible for monitoring and controlling the incoming and outgoing network traffic transmitted from/to the Internet. 
Firewall 1 allows the incoming TCP connection from the Internet to the web server on Ports 80 and 443 and the incoming TCP connection from the Internet to the Prediction Service on port 8080. In addition, Firewall 1 allows the outgoing TCP connection from the LAN to the Internet on port 80. 
All other types of communication are blocked.
    
\item \textbf{Firewall 2.} This component is responsible for monitoring and controlling the incoming and outgoing network traffic transmitted from/to the Internet or DMZ. 
Firewall 2 allows the incoming TCP connection from the web server to the feature store on port 3306 and the incoming TCP connection from the prediction service to the feature store on port 3306. 
All other types of communication are blocked.
\end{enumerate}
    
\subsection{Attack Graph Analysis}
\begin{figure}[h]
            \centering \includegraphics[width=0.95\textwidth]{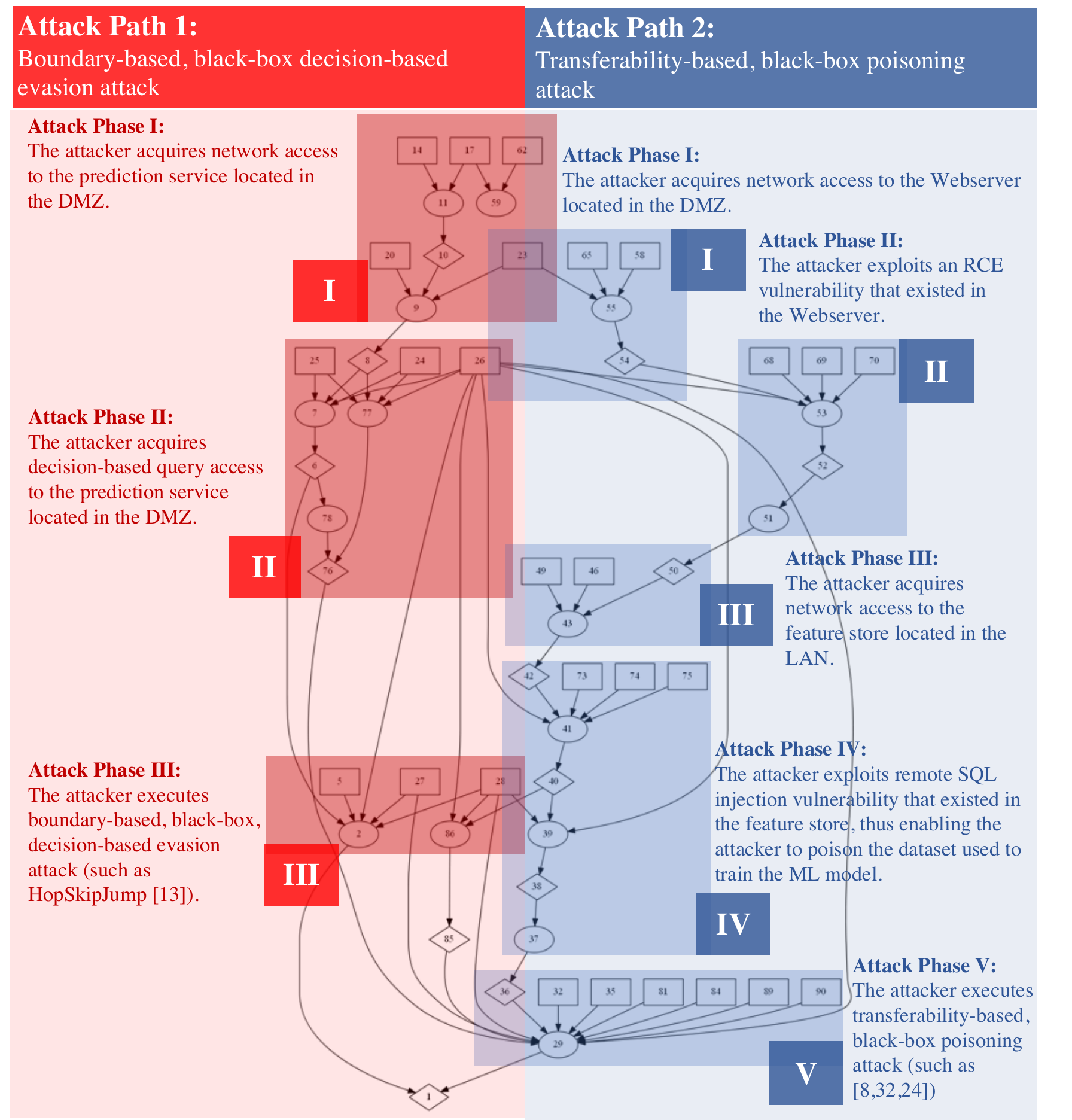}
             \caption{The attack graph generated by using the proposed extension.}
            \label{fig:attack_graph}
\end{figure}

We now use the proposed extension to generate an attack graph for the target environment. To produce an attack graph that is not too large (which will make the demonstration difficult to follow), we made the following two decisions:
First, we limit the attack graph analysis framework to evaluate the materialization of just the first threat \textit{[T1]}, i.e., evading an ML system. 
Second, we limit the adversarial attack techniques considered in the assessment to boundary-based, black-box, decision-based, evasion attacks \textit{[AT3]} and transferability-based, targeted, black-box, poisoning attacks \textit{[AT7]}. 
The input and interaction rule files are provided as supplementary material.
The resulting attack graph is presented in Figure~\ref{fig:attack_graph}.

As can be seen, the attack graph generated includes the following two main attack paths:

\noindent \textbf{Attack Path 1.} This attack path (in red) materializes the evasion threat by implementing a boundary-based, black-box, decision-based, evasion attack. This attack path, which includes three phases, is based solely on adversarial ML. In the first phase, the attacker acquires network access to the prediction service located in the DMZ. 
This is made possible because the prediction service is reachable from the Internet on port 3036. In the second phase, the attacker acquires decision-based query access in the ML pipeline. 
This is made possible, because the prediction service exposes a serving API to the client application, which is controlled by the attacker. In the third phase, the attacker executes a boundary-based, black-box, decision-based, evasion attack.
Note that this attack path is based solely on AML attack techniques.

In order to quantify the risk of this attack path we follow the procedure described in Section \ref{sec:attack_graph_risk}. 
Specifically, within attack path 1 the AML vulnerability is exploited by executing black-box evasion attack.
According to CMLVSS, the performance of this attack is classified as Medium, which resulted in a probability of 0.6. In addition, the impact of this attack is classified as tampering, which translated to an impact value of 0.33. Thus, according to  Algorithm \ref{alg:build-risk-eq}, the risk of this path are equal to 0.2.

\noindent \textbf{Attack Path 2.} This attack path (in blue) materializes the evasion threat by implementing a transferability-based, black-box, poisoning attack. This attack path, which includes five phases, combines the exploitation of traditional cybersecurity vulnerabilities as well as AML. 
In the first phase, the attacker acquires network access to the web server located in the DMZ. This is possible, because the web server is reachable from the Internet on port 80. 
In the second phase, the attacker exploits a remote code execution vulnerability that exists in the Apache web server, which results in the attacker obtaining remote control of the web server.
In the third phase, the attacker leverages this remote control and acquires network access to the feature store located in the LAN. This is possible, because the feature store listens for incoming requests on TCP port 3036, and firewall 2 allows the incoming communication from the web server to the feature store on TCP port 3036. 
In the fourth phase, the attacker exploits an SQL injection vulnerability in the feature store, which could give the attacker the ability to manipulate the data stored in the feature store. 
Since this data is used by the ML pipeline for training the learning algorithm, by manipulating the data stored in the feature store, the attacker can poison the ML model's training dataset. 
In the fifth phase, the attacker executes a transferability-based, black-box, poisoning attack, which can materialize the evasion threat. 

In order to quantify the risk of this attack path we follow the procedure described in Section \ref{sec:attack_graph_risk}. 
Specifically, in attack path 2 the attacker exploits the following three vulnerabilities:
(a) A traditional vulnerability within the Webserver. According to CVSS 2.1, this vulnerability can be exploited with LOW complexity, which translated to a probability of 0.71. 
(b) A traditional vulnerability within the SQL server.
According to CVSS 2.1, this vulnerability can be exploited with MEDIUM complexity, which translated to a probability of 0.61. 
(b) AML vulnerability which is exploited by executing transferability-based poisoning attack.
According to CMLVSS, the performance of this attack is classified as LOW, which resulted in a probability of 0.35. 
In addition, the impact of this attack is classified as tampering and denial of service, which translated to an impact value of 0.66. 
Thus, according to  Algorithm \ref{alg:build-risk-eq}, the risk of this path are equal to 0.1.
That is, although the impact of a poisoning attacks is considered more severe, the algorithm suggest to first eliminate attack path 1. 
The main reason for that is because attack path 2 require the exploitation of additional vulnerabilities to achieve the preconditions required for executing poisoning attack.

\section{\label{sec:relate_works}Related Works}

Previous works proposed various taxonomies for the security and privacy of ML systems.
For example, Papernot et al.~\cite{papernot2018sok} analyzed the security of ML systems while considering the adversary's capabilities (e.g., white-box or black-box, different threats (attack during inference phase or training phase), and adversary's goal (integrity, privacy, or availability; targeted vs. indiscriminate). 
Huang et al.~\cite{huang2011adversarial} proposed a taxonomy for classifying attacks against online ML algorithms and demonstrate attacks in two use-cases (spam detection and network anomaly detection).
The authors also considered how an adversary attacks a ML system, and discussed possible defenses and countermeasures.

In their work, Biggio et al.~\cite{biggio2018wild} provide a thorough overview of the evolution of adversarial learning research domain over a period of ten years, specifically in computer vision and cybersecurity tasks.
In their attacker model, the authors distinguish between attacker's knowledge and attacker's capabilities.

Kumar et al.~\cite{kumar2020adversarial} conducted a survey containing 28 different organizations.
Based on the results, the authors found that industry practitioners are not equipped with the proper knowledge and tools to protect, detect and respond to attacks on their ML systems.
In addition, based on the questionnaires, the authors enumerate the gaps and possible approaches to improve the security of ML systems during the development life-cycle.

Spring et al.~\cite{spring2020managing} discussed the idea of managing vulnerabilities in ML systems according to the CVE.
The authors consider only the ML algorithms and model objects.

Lin and Biggio~\cite{lin2021adversarial} mention that there is a gap between AML attacks in laboratories and the real-world attacks, however, the focus is on whether the attack can successfully transferred from a research environment to a production environment. 

The works described above are different from ours in the following aspects.
First, none of the works above consider the differences and the unique properties of production ML-based systems, which, as presented in this work, introduces additional attack vectors.
Second, none of the previous works suggest a practical method for quantifying the risk of a ML-based attack on the systems.
Finally, we provide practical and threat analysis of ML-based systems that presents a comprehensive list of attack techniques and the ability use these attack techniques within a attack graph-based threat analysis process that combines both cyber and AML threats.

\section{Conclusions}
We conducted a comprehensive threat analysis of ML production systems. 
Specifically, we pointed out the differences between ML research and production pipelines. 
We identified the primary assets of ML production systems and enumerated the threats to these assets. 
We presented an extensive threat model, which considers the attacker's access capabilities, knowledge, and goal (this threat model integrates multiple threat models described in previous works). 
We reviewed various attack techniques against ML systems.
For each attack technique, we provided a brief description of the attack method, identified the assets targeted by the adversary, and mentioned the relevant threat model.
In addition to conducting a comprehensive threat analysis of ML production systems, we developed an extension to the MulVAL attack-graph analysis framework that incorporates cyberattacks on ML production systems. 
The proposed extension enables security practitioners to apply attack graph analysis methods in environments that include ML components, thus providing security experts with a practical tool for evaluating the impact and quantifying the risk of a cyberattack targeting an ML production system.

\bibliographystyle{ACM-Reference-Format}
\bibliography{references}


\begin{thebibliography}{54}

%%% the Web address.

\ifx \showCODEN    \undefined \def \showCODEN     #1{\unskip}     \fi
\ifx \showDOI      \undefined \def \showDOI       #1{#1}\fi
\ifx \showISBNx    \undefined \def \showISBNx     #1{\unskip}     \fi
\ifx \showISBNxiii \undefined \def \showISBNxiii  #1{\unskip}     \fi
\ifx \showISSN     \undefined \def \showISSN      #1{\unskip}     \fi
\ifx \showLCCN     \undefined \def \showLCCN      #1{\unskip}     \fi
\ifx \shownote     \undefined \def \shownote      #1{#1}          \fi
\ifx \showarticletitle \undefined \def \showarticletitle #1{#1}   \fi
\ifx \showURL      \undefined \def \showURL       {\relax}        \fi
% invisible to TeX
\providecommand\bibfield[2]{#2}
\providecommand\bibinfo[2]{#2}
\providecommand\natexlab[1]{#1}
\providecommand\showeprint[2][]{arXiv:#2}

\bibitem[\protect\citeauthoryear{??}{nes}{[n.\,d.]}]%
        {nessus}
 \bibinfo{year}{[n.\,d.]}\natexlab{}.
\newblock \bibinfo{title}{{Nessus security scanner}}.
\newblock \bibinfo{howpublished}{\url{http://www.nessus.org}}.
\newblock
\newblock
\shownote{[Online]}.


\bibitem[\protect\citeauthoryear{??}{nvd}{[n.\,d.]}]%
        {nvd}
 \bibinfo{year}{[n.\,d.]}\natexlab{}.
\newblock \bibinfo{title}{{NVD national vulnerability database}}.
\newblock \bibinfo{howpublished}{\url{http://www.nvd.nist.gov}}.
\newblock
\newblock
\shownote{[Online]}.


\bibitem[\protect\citeauthoryear{Acosta, Padilla, and Homer}{Acosta
  et~al\mbox{.}}{2016}]%
        {acosta2016augmenting}
\bibfield{author}{\bibinfo{person}{Jaime~C Acosta}, \bibinfo{person}{Edgar
  Padilla}, {and} \bibinfo{person}{John Homer}.}
  \bibinfo{year}{2016}\natexlab{}.
\newblock \showarticletitle{Augmenting attack graphs to represent data link and
  network layer vulnerabilities}. In \bibinfo{booktitle}{\emph{Military
  Communications Conference, MILCOM 2016-2016 IEEE}}. IEEE,
  \bibinfo{pages}{1010--1015}.
\newblock


\bibitem[\protect\citeauthoryear{Agmon, Shabtai, and Puzis}{Agmon
  et~al\mbox{.}}{2019}]%
        {agmon2019deployment}
\bibfield{author}{\bibinfo{person}{Noga Agmon}, \bibinfo{person}{Asaf Shabtai},
  {and} \bibinfo{person}{Rami Puzis}.} \bibinfo{year}{2019}\natexlab{}.
\newblock \showarticletitle{Deployment optimization of IoT devices through
  attack graph analysis}. In \bibinfo{booktitle}{\emph{Proceedings of the 12th
  Conference on Security and Privacy in Wireless and Mobile Networks}}.
  \bibinfo{pages}{192--202}.
\newblock


\bibitem[\protect\citeauthoryear{Ateniese, Mancini, Spognardi, Villani, Vitali,
  and Felici}{Ateniese et~al\mbox{.}}{2015}]%
        {ateniese2015hacking}
\bibfield{author}{\bibinfo{person}{Giuseppe Ateniese}, \bibinfo{person}{Luigi~V
  Mancini}, \bibinfo{person}{Angelo Spognardi}, \bibinfo{person}{Antonio
  Villani}, \bibinfo{person}{Domenico Vitali}, {and} \bibinfo{person}{Giovanni
  Felici}.} \bibinfo{year}{2015}\natexlab{}.
\newblock \showarticletitle{Hacking smart machines with smarter ones: How to
  extract meaningful data from machine learning classifiers}.
\newblock \bibinfo{journal}{\emph{International Journal of Security and
  Networks}} \bibinfo{volume}{10}, \bibinfo{number}{3} (\bibinfo{year}{2015}),
  \bibinfo{pages}{137--150}.
\newblock


\bibitem[\protect\citeauthoryear{Bacic, Froh, and Henderson}{Bacic
  et~al\mbox{.}}{2006}]%
        {bacic2006mulval}
\bibfield{author}{\bibinfo{person}{Eugen Bacic}, \bibinfo{person}{Michael
  Froh}, {and} \bibinfo{person}{Glen Henderson}.}
  \bibinfo{year}{2006}\natexlab{}.
\newblock \bibinfo{booktitle}{\emph{Mulval extensions for dynamic asset
  protection}}.
\newblock \bibinfo{type}{{T}echnical {R}eport}. \bibinfo{institution}{CINNABAR
  NETWORKS INC OTTAWA (ONTARIO)}.
\newblock


\bibitem[\protect\citeauthoryear{Baylor, Breck, Cheng, Fiedel, Foo, Haque,
  Haykal, Ispir, Jain, Koc, et~al\mbox{.}}{Baylor et~al\mbox{.}}{2017}]%
        {baylor2017tfx}
\bibfield{author}{\bibinfo{person}{Denis Baylor}, \bibinfo{person}{Eric Breck},
  \bibinfo{person}{Heng-Tze Cheng}, \bibinfo{person}{Noah Fiedel},
  \bibinfo{person}{Chuan~Yu Foo}, \bibinfo{person}{Zakaria Haque},
  \bibinfo{person}{Salem Haykal}, \bibinfo{person}{Mustafa Ispir},
  \bibinfo{person}{Vihan Jain}, \bibinfo{person}{Levent Koc}, {et~al\mbox{.}}}
  \bibinfo{year}{2017}\natexlab{}.
\newblock \showarticletitle{Tfx: A tensorflow-based production-scale machine
  learning platform}. In \bibinfo{booktitle}{\emph{Proceedings of the 23rd ACM
  SIGKDD International Conference on Knowledge Discovery and Data Mining}}.
  \bibinfo{pages}{1387--1395}.
\newblock


\bibitem[\protect\citeauthoryear{Biggio, Nelson, and Laskov}{Biggio
  et~al\mbox{.}}{2012}]%
        {biggio2012poisoning}
\bibfield{author}{\bibinfo{person}{Battista Biggio}, \bibinfo{person}{Blaine
  Nelson}, {and} \bibinfo{person}{Pavel Laskov}.}
  \bibinfo{year}{2012}\natexlab{}.
\newblock \showarticletitle{Poisoning attacks against support vector machines}.
\newblock \bibinfo{journal}{\emph{arXiv preprint arXiv:1206.6389}}
  (\bibinfo{year}{2012}).
\newblock


\bibitem[\protect\citeauthoryear{Biggio and Roli}{Biggio and Roli}{2018}]%
        {biggio2018wild}
\bibfield{author}{\bibinfo{person}{Battista Biggio} {and}
  \bibinfo{person}{Fabio Roli}.} \bibinfo{year}{2018}\natexlab{}.
\newblock \showarticletitle{Wild patterns: Ten years after the rise of
  adversarial machine learning}.
\newblock \bibinfo{journal}{\emph{Pattern Recognition}}  \bibinfo{volume}{84}
  (\bibinfo{year}{2018}), \bibinfo{pages}{317--331}.
\newblock


\bibitem[\protect\citeauthoryear{Binyamini, Bitton, Inokuchi, Yagyu, Elovici,
  and Shabtai}{Binyamini et~al\mbox{.}}{2020}]%
        {binyamini2020automated}
\bibfield{author}{\bibinfo{person}{Hodaya Binyamini}, \bibinfo{person}{Ron
  Bitton}, \bibinfo{person}{Masaki Inokuchi}, \bibinfo{person}{Tomohiko Yagyu},
  \bibinfo{person}{Yuval Elovici}, {and} \bibinfo{person}{Asaf Shabtai}.}
  \bibinfo{year}{2020}\natexlab{}.
\newblock \showarticletitle{An Automated, End-to-End Framework for Modeling
  Attacks From Vulnerability Descriptions}.
\newblock \bibinfo{journal}{\emph{arXiv preprint arXiv:2008.04377}}
  (\bibinfo{year}{2020}).
\newblock


\bibitem[\protect\citeauthoryear{Carlini and Wagner}{Carlini and
  Wagner}{2017}]%
        {carlini2017adversarial}
\bibfield{author}{\bibinfo{person}{Nicholas Carlini} {and}
  \bibinfo{person}{David Wagner}.} \bibinfo{year}{2017}\natexlab{}.
\newblock \showarticletitle{Adversarial examples are not easily detected:
  Bypassing ten detection methods}. In \bibinfo{booktitle}{\emph{Proceedings of
  the 10th ACM Workshop on Artificial Intelligence and Security}}.
  \bibinfo{pages}{3--14}.
\newblock


\bibitem[\protect\citeauthoryear{Chandrasekaran, Chaudhuri, Giacomelli, Jha,
  and Yan}{Chandrasekaran et~al\mbox{.}}{2020}]%
        {chandrasekaran2020exploring}
\bibfield{author}{\bibinfo{person}{Varun Chandrasekaran},
  \bibinfo{person}{Kamalika Chaudhuri}, \bibinfo{person}{Irene Giacomelli},
  \bibinfo{person}{Somesh Jha}, {and} \bibinfo{person}{Songbai Yan}.}
  \bibinfo{year}{2020}\natexlab{}.
\newblock \showarticletitle{Exploring connections between active learning and
  model extraction}. In \bibinfo{booktitle}{\emph{29th $\{$USENIX$\}$ Security
  Symposium ($\{$USENIX$\}$ Security 20)}}. \bibinfo{pages}{1309--1326}.
\newblock


\bibitem[\protect\citeauthoryear{Chen, Jordan, and Wainwright}{Chen
  et~al\mbox{.}}{2020}]%
        {chen2020hopskipjumpattack}
\bibfield{author}{\bibinfo{person}{Jianbo Chen}, \bibinfo{person}{Michael~I
  Jordan}, {and} \bibinfo{person}{Martin~J Wainwright}.}
  \bibinfo{year}{2020}\natexlab{}.
\newblock \showarticletitle{Hopskipjumpattack: A query-efficient decision-based
  attack}. In \bibinfo{booktitle}{\emph{2020 ieee symposium on security and
  privacy (sp)}}. IEEE, \bibinfo{pages}{1277--1294}.
\newblock


\bibitem[\protect\citeauthoryear{Chen, Zhang, Sharma, Yi, and Hsieh}{Chen
  et~al\mbox{.}}{2017}]%
        {chen2017zoo}
\bibfield{author}{\bibinfo{person}{Pin-Yu Chen}, \bibinfo{person}{Huan Zhang},
  \bibinfo{person}{Yash Sharma}, \bibinfo{person}{Jinfeng Yi}, {and}
  \bibinfo{person}{Cho-Jui Hsieh}.} \bibinfo{year}{2017}\natexlab{}.
\newblock \showarticletitle{Zoo: Zeroth order optimization based black-box
  attacks to deep neural networks without training substitute models}. In
  \bibinfo{booktitle}{\emph{Proceedings of the 10th ACM workshop on artificial
  intelligence and security}}. \bibinfo{pages}{15--26}.
\newblock


\bibitem[\protect\citeauthoryear{Downing}{Downing}{2019}]%
        {downing2019mlsploit}
\bibfield{author}{\bibinfo{person}{Evan Downing}.}
  \bibinfo{year}{2019}\natexlab{}.
\newblock \showarticletitle{MLsploit [Judges Remarks]}.
\newblock  (\bibinfo{year}{2019}).
\newblock


\bibitem[\protect\citeauthoryear{El~Mir, Kandoussi, Hanini, Haqiq, and
  Kim}{El~Mir et~al\mbox{.}}{2017}]%
        {el2017game}
\bibfield{author}{\bibinfo{person}{Iman El~Mir}, \bibinfo{person}{El~Mehdi
  Kandoussi}, \bibinfo{person}{Mohamed Hanini}, \bibinfo{person}{Abdelkrim
  Haqiq}, {and} \bibinfo{person}{Dong~Seong Kim}.}
  \bibinfo{year}{2017}\natexlab{}.
\newblock \showarticletitle{A Game Theoretic approach based virtual machine
  migration for cloud environment security}.
\newblock \bibinfo{journal}{\emph{International Journal of Communication
  Networks and Information Security}} \bibinfo{volume}{9}, \bibinfo{number}{3}
  (\bibinfo{year}{2017}), \bibinfo{pages}{345--357}.
\newblock


\bibitem[\protect\citeauthoryear{Fredrikson, Jha, and Ristenpart}{Fredrikson
  et~al\mbox{.}}{2015}]%
        {fredrikson2015model}
\bibfield{author}{\bibinfo{person}{Matt Fredrikson}, \bibinfo{person}{Somesh
  Jha}, {and} \bibinfo{person}{Thomas Ristenpart}.}
  \bibinfo{year}{2015}\natexlab{}.
\newblock \showarticletitle{Model inversion attacks that exploit confidence
  information and basic countermeasures}. In
  \bibinfo{booktitle}{\emph{Proceedings of the 22nd ACM SIGSAC Conference on
  Computer and Communications Security}}. \bibinfo{pages}{1322--1333}.
\newblock


\bibitem[\protect\citeauthoryear{Fredrikson, Lantz, Jha, Lin, Page, and
  Ristenpart}{Fredrikson et~al\mbox{.}}{2014}]%
        {fredrikson2014privacy}
\bibfield{author}{\bibinfo{person}{Matthew Fredrikson}, \bibinfo{person}{Eric
  Lantz}, \bibinfo{person}{Somesh Jha}, \bibinfo{person}{Simon Lin},
  \bibinfo{person}{David Page}, {and} \bibinfo{person}{Thomas Ristenpart}.}
  \bibinfo{year}{2014}\natexlab{}.
\newblock \showarticletitle{Privacy in pharmacogenetics: An end-to-end case
  study of personalized warfarin dosing}. In \bibinfo{booktitle}{\emph{23rd
  $\{$USENIX$\}$ Security Symposium ($\{$USENIX$\}$ Security 14)}}.
  \bibinfo{pages}{17--32}.
\newblock


\bibitem[\protect\citeauthoryear{Froh and Henderson}{Froh and
  Henderson}{2009}]%
        {froh2009mulval}
\bibfield{author}{\bibinfo{person}{Michael~John Froh} {and}
  \bibinfo{person}{Glen Henderson}.} \bibinfo{year}{2009}\natexlab{}.
\newblock \bibinfo{booktitle}{\emph{MulVAL extensions II}}.
\newblock \bibinfo{publisher}{Defence R \& D Canada-Ottawa}.
\newblock


\bibitem[\protect\citeauthoryear{Ganju, Wang, Yang, Gunter, and Borisov}{Ganju
  et~al\mbox{.}}{2018}]%
        {ganju2018property}
\bibfield{author}{\bibinfo{person}{Karan Ganju}, \bibinfo{person}{Qi Wang},
  \bibinfo{person}{Wei Yang}, \bibinfo{person}{Carl~A Gunter}, {and}
  \bibinfo{person}{Nikita Borisov}.} \bibinfo{year}{2018}\natexlab{}.
\newblock \showarticletitle{Property inference attacks on fully connected
  neural networks using permutation invariant representations}. In
  \bibinfo{booktitle}{\emph{Proceedings of the 2018 ACM SIGSAC Conference on
  Computer and Communications Security}}. \bibinfo{pages}{619--633}.
\newblock


\bibitem[\protect\citeauthoryear{Gonda, Pascal, Puzis, Shani, and
  Shapira}{Gonda et~al\mbox{.}}{2018}]%
        {gonda2018analysis}
\bibfield{author}{\bibinfo{person}{Tom Gonda}, \bibinfo{person}{Tal Pascal},
  \bibinfo{person}{Rami Puzis}, \bibinfo{person}{Guy Shani}, {and}
  \bibinfo{person}{Bracha Shapira}.} \bibinfo{year}{2018}\natexlab{}.
\newblock \showarticletitle{Analysis of Attack Graph Representations for
  Ranking Vulnerability Fixes.}. In \bibinfo{booktitle}{\emph{GCAI}}.
  \bibinfo{pages}{215--228}.
\newblock


\bibitem[\protect\citeauthoryear{Goodfellow, Shlens, and Szegedy}{Goodfellow
  et~al\mbox{.}}{2014}]%
        {goodfellow2014explaining}
\bibfield{author}{\bibinfo{person}{Ian~J Goodfellow}, \bibinfo{person}{Jonathon
  Shlens}, {and} \bibinfo{person}{Christian Szegedy}.}
  \bibinfo{year}{2014}\natexlab{}.
\newblock \showarticletitle{Explaining and harnessing adversarial examples}.
\newblock \bibinfo{journal}{\emph{arXiv preprint arXiv:1412.6572}}
  (\bibinfo{year}{2014}).
\newblock


\bibitem[\protect\citeauthoryear{Hidano, Murakami, Katsumata, Kiyomoto, and
  Hanaoka}{Hidano et~al\mbox{.}}{2017}]%
        {hidano2017model}
\bibfield{author}{\bibinfo{person}{Seira Hidano}, \bibinfo{person}{Takao
  Murakami}, \bibinfo{person}{Shuichi Katsumata}, \bibinfo{person}{Shinsaku
  Kiyomoto}, {and} \bibinfo{person}{Goichiro Hanaoka}.}
  \bibinfo{year}{2017}\natexlab{}.
\newblock \showarticletitle{Model inversion attacks for prediction systems:
  Without knowledge of non-sensitive attributes}. In
  \bibinfo{booktitle}{\emph{2017 15th Annual Conference on Privacy, Security
  and Trust (PST)}}. IEEE, \bibinfo{pages}{115--11509}.
\newblock


\bibitem[\protect\citeauthoryear{Huang, Joseph, Nelson, Rubinstein, and
  Tygar}{Huang et~al\mbox{.}}{2011}]%
        {huang2011adversarial}
\bibfield{author}{\bibinfo{person}{Ling Huang}, \bibinfo{person}{Anthony~D
  Joseph}, \bibinfo{person}{Blaine Nelson}, \bibinfo{person}{Benjamin~IP
  Rubinstein}, {and} \bibinfo{person}{J~Doug Tygar}.}
  \bibinfo{year}{2011}\natexlab{}.
\newblock \showarticletitle{Adversarial machine learning}. In
  \bibinfo{booktitle}{\emph{Proceedings of the 4th ACM workshop on Security and
  artificial intelligence}}. \bibinfo{pages}{43--58}.
\newblock


\bibitem[\protect\citeauthoryear{Inokuchi, Ohta, Kinoshita, Yagyu, Stan,
  Bitton, Elovici, and Shabtai}{Inokuchi et~al\mbox{.}}{2019}]%
        {inokuchi2019design}
\bibfield{author}{\bibinfo{person}{Masaki Inokuchi}, \bibinfo{person}{Yoshinobu
  Ohta}, \bibinfo{person}{Shunichi Kinoshita}, \bibinfo{person}{Tomohiko
  Yagyu}, \bibinfo{person}{Orly Stan}, \bibinfo{person}{Ron Bitton},
  \bibinfo{person}{Yuval Elovici}, {and} \bibinfo{person}{Asaf Shabtai}.}
  \bibinfo{year}{2019}\natexlab{}.
\newblock \showarticletitle{Design Procedure of Knowledge Base for Practical
  Attack Graph Generation}. In \bibinfo{booktitle}{\emph{Proceedings of the
  2019 ACM Asia Conference on Computer and Communications Security}}.
  \bibinfo{pages}{594--601}.
\newblock


\bibitem[\protect\citeauthoryear{Jagielski, Oprea, Biggio, Liu, Nita-Rotaru,
  and Li}{Jagielski et~al\mbox{.}}{2018}]%
        {jagielski2018manipulating}
\bibfield{author}{\bibinfo{person}{Matthew Jagielski}, \bibinfo{person}{Alina
  Oprea}, \bibinfo{person}{Battista Biggio}, \bibinfo{person}{Chang Liu},
  \bibinfo{person}{Cristina Nita-Rotaru}, {and} \bibinfo{person}{Bo Li}.}
  \bibinfo{year}{2018}\natexlab{}.
\newblock \showarticletitle{Manipulating machine learning: Poisoning attacks
  and countermeasures for regression learning}. In
  \bibinfo{booktitle}{\emph{2018 IEEE Symposium on Security and Privacy (SP)}}.
  IEEE, \bibinfo{pages}{19--35}.
\newblock


\bibitem[\protect\citeauthoryear{Juuti, Szyller, Marchal, and Asokan}{Juuti
  et~al\mbox{.}}{2019}]%
        {juuti2019prada}
\bibfield{author}{\bibinfo{person}{Mika Juuti}, \bibinfo{person}{Sebastian
  Szyller}, \bibinfo{person}{Samuel Marchal}, {and} \bibinfo{person}{N
  Asokan}.} \bibinfo{year}{2019}\natexlab{}.
\newblock \showarticletitle{PRADA: protecting against DNN model stealing
  attacks}. In \bibinfo{booktitle}{\emph{2019 IEEE European Symposium on
  Security and Privacy (EuroS\&P)}}. IEEE, \bibinfo{pages}{512--527}.
\newblock


\bibitem[\protect\citeauthoryear{Kravchik, Biggio, and Shabtai}{Kravchik
  et~al\mbox{.}}{2021}]%
        {kravchik2021poisoning}
\bibfield{author}{\bibinfo{person}{Moshe Kravchik}, \bibinfo{person}{Battista
  Biggio}, {and} \bibinfo{person}{Asaf Shabtai}.}
  \bibinfo{year}{2021}\natexlab{}.
\newblock \showarticletitle{Poisoning attacks on cyber attack detectors for
  industrial control systems}. In \bibinfo{booktitle}{\emph{Proceedings of the
  36th Annual ACM Symposium on Applied Computing}}. \bibinfo{pages}{116--125}.
\newblock


\bibitem[\protect\citeauthoryear{Kumar, Nystr{\"o}m, Lambert, Marshall,
  Goertzel, Comissoneru, Swann, and Xia}{Kumar et~al\mbox{.}}{2020}]%
        {kumar2020adversarial}
\bibfield{author}{\bibinfo{person}{Ram Shankar~Siva Kumar},
  \bibinfo{person}{Magnus Nystr{\"o}m}, \bibinfo{person}{John Lambert},
  \bibinfo{person}{Andrew Marshall}, \bibinfo{person}{Mario Goertzel},
  \bibinfo{person}{Andi Comissoneru}, \bibinfo{person}{Matt Swann}, {and}
  \bibinfo{person}{Sharon Xia}.} \bibinfo{year}{2020}\natexlab{}.
\newblock \showarticletitle{Adversarial machine learning-industry
  perspectives}. In \bibinfo{booktitle}{\emph{2020 IEEE Security and Privacy
  Workshops (SPW)}}. IEEE, \bibinfo{pages}{69--75}.
\newblock


\bibitem[\protect\citeauthoryear{Kurakin, Goodfellow, Bengio,
  et~al\mbox{.}}{Kurakin et~al\mbox{.}}{2016}]%
        {kurakin2016adversarial}
\bibfield{author}{\bibinfo{person}{Alexey Kurakin}, \bibinfo{person}{Ian
  Goodfellow}, \bibinfo{person}{Samy Bengio}, {et~al\mbox{.}}}
  \bibinfo{year}{2016}\natexlab{}.
\newblock \bibinfo{title}{Adversarial examples in the physical world}.
\newblock
\newblock


\bibitem[\protect\citeauthoryear{Lin and Biggio}{Lin and Biggio}{2021}]%
        {lin2021adversarial}
\bibfield{author}{\bibinfo{person}{Hsiao-Ying Lin} {and}
  \bibinfo{person}{Battista Biggio}.} \bibinfo{year}{2021}\natexlab{}.
\newblock \showarticletitle{Adversarial Machine Learning: Attacks From
  Laboratories to the Real World}.
\newblock \bibinfo{journal}{\emph{Computer}} \bibinfo{volume}{54},
  \bibinfo{number}{5} (\bibinfo{year}{2021}), \bibinfo{pages}{56--60}.
\newblock


\bibitem[\protect\citeauthoryear{Liu, Singhal, and Wijesekera}{Liu
  et~al\mbox{.}}{2015}]%
        {liu2015logic}
\bibfield{author}{\bibinfo{person}{Changwei Liu}, \bibinfo{person}{Anoop
  Singhal}, {and} \bibinfo{person}{Duminda Wijesekera}.}
  \bibinfo{year}{2015}\natexlab{}.
\newblock \showarticletitle{A logic-based network forensic model for evidence
  analysis}. In \bibinfo{booktitle}{\emph{IFIP International Conference on
  Digital Forensics}}. Springer, \bibinfo{pages}{129--145}.
\newblock


\bibitem[\protect\citeauthoryear{Mavani and Asawa}{Mavani and Asawa}{2017}]%
        {mavani2017modeling}
\bibfield{author}{\bibinfo{person}{Monali Mavani} {and}
  \bibinfo{person}{Krishna Asawa}.} \bibinfo{year}{2017}\natexlab{}.
\newblock \showarticletitle{Modeling and analyses of IP spoofing attack in
  6LoWPAN network}.
\newblock \bibinfo{journal}{\emph{Computers \& Security}}  \bibinfo{volume}{70}
  (\bibinfo{year}{2017}), \bibinfo{pages}{95--110}.
\newblock


\bibitem[\protect\citeauthoryear{Mei and Zhu}{Mei and Zhu}{2015}]%
        {mei2015using}
\bibfield{author}{\bibinfo{person}{Shike Mei} {and} \bibinfo{person}{Xiaojin
  Zhu}.} \bibinfo{year}{2015}\natexlab{}.
\newblock \showarticletitle{Using machine teaching to identify optimal
  training-set attacks on machine learners}. In
  \bibinfo{booktitle}{\emph{Proceedings of the AAAI Conference on Artificial
  Intelligence}}, Vol.~\bibinfo{volume}{29}.
\newblock


\bibitem[\protect\citeauthoryear{Melis, Demontis, Pintor, Sotgiu, and
  Biggio}{Melis et~al\mbox{.}}{2019}]%
        {melis2019secml}
\bibfield{author}{\bibinfo{person}{Marco Melis}, \bibinfo{person}{Ambra
  Demontis}, \bibinfo{person}{Maura Pintor}, \bibinfo{person}{Angelo Sotgiu},
  {and} \bibinfo{person}{Battista Biggio}.} \bibinfo{year}{2019}\natexlab{}.
\newblock \showarticletitle{secml: A python library for secure and explainable
  machine learning}.
\newblock \bibinfo{journal}{\emph{arXiv preprint arXiv:1912.10013}}
  (\bibinfo{year}{2019}).
\newblock


\bibitem[\protect\citeauthoryear{Mu{\~n}oz-Gonz{\'a}lez, Biggio, Demontis,
  Paudice, Wongrassamee, Lupu, and Roli}{Mu{\~n}oz-Gonz{\'a}lez
  et~al\mbox{.}}{2017}]%
        {munoz2017towards}
\bibfield{author}{\bibinfo{person}{Luis Mu{\~n}oz-Gonz{\'a}lez},
  \bibinfo{person}{Battista Biggio}, \bibinfo{person}{Ambra Demontis},
  \bibinfo{person}{Andrea Paudice}, \bibinfo{person}{Vasin Wongrassamee},
  \bibinfo{person}{Emil~C Lupu}, {and} \bibinfo{person}{Fabio Roli}.}
  \bibinfo{year}{2017}\natexlab{}.
\newblock \showarticletitle{Towards poisoning of deep learning algorithms with
  back-gradient optimization}. In \bibinfo{booktitle}{\emph{Proceedings of the
  10th ACM Workshop on Artificial Intelligence and Security}}.
  \bibinfo{pages}{27--38}.
\newblock


\bibitem[\protect\citeauthoryear{Nasr, Shokri, and Houmansadr}{Nasr
  et~al\mbox{.}}{2019}]%
        {nasr2019comprehensive}
\bibfield{author}{\bibinfo{person}{Milad Nasr}, \bibinfo{person}{Reza Shokri},
  {and} \bibinfo{person}{Amir Houmansadr}.} \bibinfo{year}{2019}\natexlab{}.
\newblock \showarticletitle{Comprehensive privacy analysis of deep learning:
  Passive and active white-box inference attacks against centralized and
  federated learning}. In \bibinfo{booktitle}{\emph{2019 IEEE symposium on
  security and privacy (SP)}}. IEEE, \bibinfo{pages}{739--753}.
\newblock


\bibitem[\protect\citeauthoryear{Nicolae, Sinn, Tran, Buesser, Rawat, Wistuba,
  Zantedeschi, Baracaldo, Chen, Ludwig, et~al\mbox{.}}{Nicolae
  et~al\mbox{.}}{2018}]%
        {nicolae2018adversarial}
\bibfield{author}{\bibinfo{person}{Maria-Irina Nicolae},
  \bibinfo{person}{Mathieu Sinn}, \bibinfo{person}{Minh~Ngoc Tran},
  \bibinfo{person}{Beat Buesser}, \bibinfo{person}{Ambrish Rawat},
  \bibinfo{person}{Martin Wistuba}, \bibinfo{person}{Valentina Zantedeschi},
  \bibinfo{person}{Nathalie Baracaldo}, \bibinfo{person}{Bryant Chen},
  \bibinfo{person}{Heiko Ludwig}, {et~al\mbox{.}}}
  \bibinfo{year}{2018}\natexlab{}.
\newblock \showarticletitle{Adversarial Robustness Toolbox v1. 0.0}.
\newblock \bibinfo{journal}{\emph{arXiv preprint arXiv:1807.01069}}
  (\bibinfo{year}{2018}).
\newblock


\bibitem[\protect\citeauthoryear{Oh, Schiele, and Fritz}{Oh
  et~al\mbox{.}}{2019}]%
        {oh2019towards}
\bibfield{author}{\bibinfo{person}{Seong~Joon Oh}, \bibinfo{person}{Bernt
  Schiele}, {and} \bibinfo{person}{Mario Fritz}.}
  \bibinfo{year}{2019}\natexlab{}.
\newblock \showarticletitle{Towards reverse-engineering black-box neural
  networks}.
\newblock In \bibinfo{booktitle}{\emph{Explainable AI: Interpreting, Explaining
  and Visualizing Deep Learning}}. \bibinfo{publisher}{Springer},
  \bibinfo{pages}{121--144}.
\newblock


\bibitem[\protect\citeauthoryear{Ou, Govindavajhala, and Appel}{Ou
  et~al\mbox{.}}{2005}]%
        {ou2005mulval}
\bibfield{author}{\bibinfo{person}{Xinming Ou}, \bibinfo{person}{Sudhakar
  Govindavajhala}, {and} \bibinfo{person}{Andrew~W Appel}.}
  \bibinfo{year}{2005}\natexlab{}.
\newblock \showarticletitle{MulVAL: A Logic-based Network Security Analyzer.}.
  In \bibinfo{booktitle}{\emph{USENIX security symposium}},
  Vol.~\bibinfo{volume}{8}. Baltimore, MD, \bibinfo{pages}{113--128}.
\newblock


\bibitem[\protect\citeauthoryear{Papernot, Faghri, Carlini, Goodfellow,
  Feinman, Kurakin, Xie, Sharma, Brown, Roy, Matyasko, Behzadan, Hambardzumyan,
  Zhang, Juang, Li, Sheatsley, Garg, Uesato, Gierke, Dong, Berthelot,
  Hendricks, Rauber, and Long}{Papernot et~al\mbox{.}}{2018a}]%
        {papernot2018cleverhans}
\bibfield{author}{\bibinfo{person}{Nicolas Papernot}, \bibinfo{person}{Fartash
  Faghri}, \bibinfo{person}{Nicholas Carlini}, \bibinfo{person}{Ian
  Goodfellow}, \bibinfo{person}{Reuben Feinman}, \bibinfo{person}{Alexey
  Kurakin}, \bibinfo{person}{Cihang Xie}, \bibinfo{person}{Yash Sharma},
  \bibinfo{person}{Tom Brown}, \bibinfo{person}{Aurko Roy},
  \bibinfo{person}{Alexander Matyasko}, \bibinfo{person}{Vahid Behzadan},
  \bibinfo{person}{Karen Hambardzumyan}, \bibinfo{person}{Zhishuai Zhang},
  \bibinfo{person}{Yi-Lin Juang}, \bibinfo{person}{Zhi Li},
  \bibinfo{person}{Ryan Sheatsley}, \bibinfo{person}{Abhibhav Garg},
  \bibinfo{person}{Jonathan Uesato}, \bibinfo{person}{Willi Gierke},
  \bibinfo{person}{Yinpeng Dong}, \bibinfo{person}{David Berthelot},
  \bibinfo{person}{Paul Hendricks}, \bibinfo{person}{Jonas Rauber}, {and}
  \bibinfo{person}{Rujun Long}.} \bibinfo{year}{2018}\natexlab{a}.
\newblock \showarticletitle{Technical Report on the CleverHans v2.1.0
  Adversarial Examples Library}.
\newblock \bibinfo{journal}{\emph{arXiv preprint arXiv:1610.00768}}
  (\bibinfo{year}{2018}).
\newblock


\bibitem[\protect\citeauthoryear{Papernot, McDaniel, and Goodfellow}{Papernot
  et~al\mbox{.}}{2016}]%
        {papernot2016transferability}
\bibfield{author}{\bibinfo{person}{Nicolas Papernot}, \bibinfo{person}{Patrick
  McDaniel}, {and} \bibinfo{person}{Ian Goodfellow}.}
  \bibinfo{year}{2016}\natexlab{}.
\newblock \showarticletitle{Transferability in machine learning: from phenomena
  to black-box attacks using adversarial samples}.
\newblock \bibinfo{journal}{\emph{arXiv preprint arXiv:1605.07277}}
  (\bibinfo{year}{2016}).
\newblock


\bibitem[\protect\citeauthoryear{Papernot, McDaniel, Goodfellow, Jha, Celik,
  and Swami}{Papernot et~al\mbox{.}}{2017}]%
        {papernot2017practical}
\bibfield{author}{\bibinfo{person}{Nicolas Papernot}, \bibinfo{person}{Patrick
  McDaniel}, \bibinfo{person}{Ian Goodfellow}, \bibinfo{person}{Somesh Jha},
  \bibinfo{person}{Z~Berkay Celik}, {and} \bibinfo{person}{Ananthram Swami}.}
  \bibinfo{year}{2017}\natexlab{}.
\newblock \showarticletitle{Practical black-box attacks against machine
  learning}. In \bibinfo{booktitle}{\emph{Proceedings of the 2017 ACM on Asia
  conference on computer and communications security}}.
  \bibinfo{pages}{506--519}.
\newblock


\bibitem[\protect\citeauthoryear{Papernot, McDaniel, Sinha, and
  Wellman}{Papernot et~al\mbox{.}}{2018b}]%
        {papernot2018sok}
\bibfield{author}{\bibinfo{person}{Nicolas Papernot}, \bibinfo{person}{Patrick
  McDaniel}, \bibinfo{person}{Arunesh Sinha}, {and} \bibinfo{person}{Michael~P
  Wellman}.} \bibinfo{year}{2018}\natexlab{b}.
\newblock \showarticletitle{Sok: Security and privacy in machine learning}. In
  \bibinfo{booktitle}{\emph{2018 IEEE European Symposium on Security and
  Privacy (EuroS\&P)}}. IEEE, \bibinfo{pages}{399--414}.
\newblock


\bibitem[\protect\citeauthoryear{Rauber, Brendel, and Bethge}{Rauber
  et~al\mbox{.}}{2017}]%
        {rauber2017foolbox}
\bibfield{author}{\bibinfo{person}{Jonas Rauber}, \bibinfo{person}{Wieland
  Brendel}, {and} \bibinfo{person}{Matthias Bethge}.}
  \bibinfo{year}{2017}\natexlab{}.
\newblock \showarticletitle{Foolbox: A python toolbox to benchmark the
  robustness of machine learning models}.
\newblock \bibinfo{journal}{\emph{arXiv preprint arXiv:1707.04131}}
  (\bibinfo{year}{2017}).
\newblock


\bibitem[\protect\citeauthoryear{Saaty}{Saaty}{2008}]%
        {saaty2008decision}
\bibfield{author}{\bibinfo{person}{Thomas~L Saaty}.}
  \bibinfo{year}{2008}\natexlab{}.
\newblock \showarticletitle{Decision making with the analytic hierarchy
  process}.
\newblock \bibinfo{journal}{\emph{International journal of services sciences}}
  \bibinfo{volume}{1}, \bibinfo{number}{1} (\bibinfo{year}{2008}),
  \bibinfo{pages}{83--98}.
\newblock


\bibitem[\protect\citeauthoryear{Shokri, Stronati, Song, and Shmatikov}{Shokri
  et~al\mbox{.}}{2017}]%
        {shokri2017membership}
\bibfield{author}{\bibinfo{person}{Reza Shokri}, \bibinfo{person}{Marco
  Stronati}, \bibinfo{person}{Congzheng Song}, {and} \bibinfo{person}{Vitaly
  Shmatikov}.} \bibinfo{year}{2017}\natexlab{}.
\newblock \showarticletitle{Membership inference attacks against machine
  learning models}. In \bibinfo{booktitle}{\emph{2017 IEEE Symposium on
  Security and Privacy (SP)}}. IEEE, \bibinfo{pages}{3--18}.
\newblock


\bibitem[\protect\citeauthoryear{Song and Raghunathan}{Song and
  Raghunathan}{2020}]%
        {song2020information}
\bibfield{author}{\bibinfo{person}{Congzheng Song} {and}
  \bibinfo{person}{Ananth Raghunathan}.} \bibinfo{year}{2020}\natexlab{}.
\newblock \showarticletitle{Information leakage in embedding models}. In
  \bibinfo{booktitle}{\emph{Proceedings of the 2020 ACM SIGSAC Conference on
  Computer and Communications Security}}. \bibinfo{pages}{377--390}.
\newblock


\bibitem[\protect\citeauthoryear{Spring, Galyardt, Householder, and
  VanHoudnos}{Spring et~al\mbox{.}}{2020}]%
        {spring2020managing}
\bibfield{author}{\bibinfo{person}{Jonathan~M Spring}, \bibinfo{person}{April
  Galyardt}, \bibinfo{person}{Allen~D Householder}, {and}
  \bibinfo{person}{Nathan VanHoudnos}.} \bibinfo{year}{2020}\natexlab{}.
\newblock \showarticletitle{On managing vulnerabilities in AI/ML systems}. In
  \bibinfo{booktitle}{\emph{New Security Paradigms Workshop 2020}}.
  \bibinfo{pages}{111--126}.
\newblock


\bibitem[\protect\citeauthoryear{Stan, Bitton, Ezrets, Dadon, Inokuchi, Ohta,
  Yagyu, Elovici, and Shabtai}{Stan et~al\mbox{.}}{[n.\,d.]}]%
        {stan2021heuristic}
\bibfield{author}{\bibinfo{person}{Orly Stan}, \bibinfo{person}{Ron Bitton},
  \bibinfo{person}{Michal Ezrets}, \bibinfo{person}{Moran Dadon},
  \bibinfo{person}{Masaki Inokuchi}, \bibinfo{person}{Yoshinobu Ohta},
  \bibinfo{person}{Tomohiko Yagyu}, \bibinfo{person}{Yuval Elovici}, {and}
  \bibinfo{person}{Asaf Shabtai}.} \bibinfo{year}{[n.\,d.]}\natexlab{}.
\newblock \showarticletitle{Heuristic Approach for Countermeasure Selection
  Using Attack Graphs}. In \bibinfo{booktitle}{\emph{2021 IEEE 34th Computer
  Security Foundations Symposium (CSF)}}. IEEE Computer Society,
  \bibinfo{pages}{63--78}.
\newblock


\bibitem[\protect\citeauthoryear{Stan, Bitton, Ezrets, Dadon, Inokuchi, Ohta,
  Yamada, Yagyu, Elovici, and Shabtai}{Stan et~al\mbox{.}}{2019}]%
        {stan2019extending}
\bibfield{author}{\bibinfo{person}{Orly Stan}, \bibinfo{person}{Ron Bitton},
  \bibinfo{person}{Michal Ezrets}, \bibinfo{person}{Moran Dadon},
  \bibinfo{person}{Masaki Inokuchi}, \bibinfo{person}{Yoshinobu Ohta},
  \bibinfo{person}{Yoshiyuki Yamada}, \bibinfo{person}{Tomohiko Yagyu},
  \bibinfo{person}{Yuval Elovici}, {and} \bibinfo{person}{Asaf Shabtai}.}
  \bibinfo{year}{2019}\natexlab{}.
\newblock \showarticletitle{Extending Attack Graphs to Represent Cyber-Attacks
  in Communication Protocols and Modern IT Networks}.
\newblock \bibinfo{journal}{\emph{arXiv preprint arXiv:1906.09786}}
  (\bibinfo{year}{2019}).
\newblock


\bibitem[\protect\citeauthoryear{Szegedy, Zaremba, Sutskever, Bruna, Erhan,
  Goodfellow, and Fergus}{Szegedy et~al\mbox{.}}{2013}]%
        {szegedy2013intriguing}
\bibfield{author}{\bibinfo{person}{Christian Szegedy},
  \bibinfo{person}{Wojciech Zaremba}, \bibinfo{person}{Ilya Sutskever},
  \bibinfo{person}{Joan Bruna}, \bibinfo{person}{Dumitru Erhan},
  \bibinfo{person}{Ian Goodfellow}, {and} \bibinfo{person}{Rob Fergus}.}
  \bibinfo{year}{2013}\natexlab{}.
\newblock \showarticletitle{Intriguing properties of neural networks}.
\newblock \bibinfo{journal}{\emph{arXiv preprint arXiv:1312.6199}}
  (\bibinfo{year}{2013}).
\newblock


\bibitem[\protect\citeauthoryear{Xiao, Biggio, Brown, Fumera, Eckert, and
  Roli}{Xiao et~al\mbox{.}}{2015}]%
        {xiao2015feature}
\bibfield{author}{\bibinfo{person}{Huang Xiao}, \bibinfo{person}{Battista
  Biggio}, \bibinfo{person}{Gavin Brown}, \bibinfo{person}{Giorgio Fumera},
  \bibinfo{person}{Claudia Eckert}, {and} \bibinfo{person}{Fabio Roli}.}
  \bibinfo{year}{2015}\natexlab{}.
\newblock \showarticletitle{Is feature selection secure against training data
  poisoning?}. In \bibinfo{booktitle}{\emph{International Conference on Machine
  Learning}}. PMLR, \bibinfo{pages}{1689--1698}.
\newblock


\bibitem[\protect\citeauthoryear{Yang, Zhang, Chang, and Liang}{Yang
  et~al\mbox{.}}{2019}]%
        {yang2019neural}
\bibfield{author}{\bibinfo{person}{Ziqi Yang}, \bibinfo{person}{Jiyi Zhang},
  \bibinfo{person}{Ee-Chien Chang}, {and} \bibinfo{person}{Zhenkai Liang}.}
  \bibinfo{year}{2019}\natexlab{}.
\newblock \showarticletitle{Neural network inversion in adversarial setting via
  background knowledge alignment}. In \bibinfo{booktitle}{\emph{Proceedings of
  the 2019 ACM SIGSAC Conference on Computer and Communications Security}}.
  \bibinfo{pages}{225--240}.
\newblock


\end{thebibliography}

\end{document}